\documentclass{ws-book961x669}
\usepackage{ws-book-har}      
\usepackage[hyperindex,pdftex]{hyperref}

\def\eq#1{{Eq.~(\ref{#1})}}
\def\fig#1{{Fig.~\ref{#1}}}
\newcommand{\as}{\alpha_s}
\newcommand{\bra}[1]{\left\langle #1 \right|}
\newcommand{\ket}[1]{\left| #1 \right\rangle}
\newcommand{\thalf}{\tfrac{1}{2}}

\title{From the past to the future --- the legacy of Lev Lipatov}         

\begin{document}

\tableofcontents

\setcounter{page}{1}

\chapter[From Parton Saturation to Proton Spin]{From Parton Saturation to Proton Spin: \\ the Impact of BFKL Equation and Reggeon Evolution}

\author{Yuri V. Kovchegov} 
\begin{center}
  {\it Department of Physics, The Ohio State University, Columbus, OH 43210, USA} \\
  Email: {\it kovchegov.1@osu.edu }
\end{center}

\begin{figure}
\begin{center}
\includegraphics[width= 0.4 \textwidth]{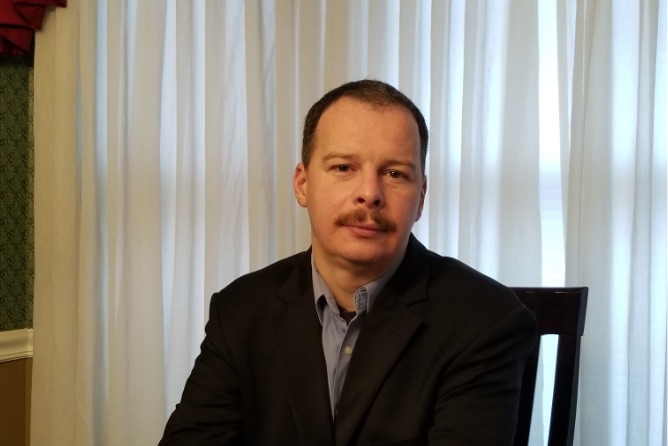} 
\end{center}
\end{figure}

\label{ch1}

Lev Lipatov was a giant in the field of strong interactions and a dominant force in high-energy QCD for many decades. His work deeply influenced both how we think about QCD and how we perform calculations in the theory. Below we describe the work in two related research directions: the physics of parton saturation and proton spin at small $x$. Both developments would have been impossible without Lipatov's groundbreaking work. Saturation physics would not have happened without the Balitsky--Fadin--Kuraev--Lipatov (BFKL) equation
\cite{Kuraev:1977fs,Balitsky:1978ic}. The recent progress in our theoretical understanding of the proton spin contribution coming from small-$x$ partons started with the seminal paper by Kirschner and Lipatov \cite{Kirschner:1983di} resumming double logarithms of energy in the Reggeon evolution (see also \cite{Gorshkov:1966ht}).




\section{Saturation and Unitarity}

\label{Sec:SU}

With the advent of QCD as the theory of strong interactions, an important question arose: what is the correct high-energy asymptotics of QCD? Lev Lipatov and collaborators tackled this question head on by deriving the BFKL equation \cite{Kuraev:1977fs,Balitsky:1978ic}, an equation describing the high-energy behavior of cross sections and parton distribution functions in QCD. The BFKL equation was a breakthrough in our understanding of QCD at high energy. The solution of BFKL equation \cite{Kuraev:1977fs,Balitsky:1978ic} leads to a cross section which grows as a small positive power of the center-of-mass energy squared $s$, 
\begin{align}
\sigma \sim s^{\mbox{const} \, \as}
\end{align}
where the constant in the exponent is positive (const~$>0$) and $\as$ is the strong coupling. This behavior violates the unitarity bound \cite{Heisenberg:1952zz,Froissart:1961ux,Martin:1969ina} 
\begin{align}
\sigma \le \ln^2 s. 
\end{align}
Therefore, the high energy QCD community understood that BFKL equation needed to be unitarized. That is, at very high energies, some corrections to BFKL evolution must become important, restoring unitarity. 

At the same time, it is known that the gluon and quark parton distribution functions (PDFs) given by the solution of BFKL evolution, or extracted from experimental data, grow as positive powers of $1/x$. While there is no unitarity bound for the PDFs, very high numbers of quarks and gluons resulting from the power-of-$1/x$ growth at small $x$ must lead to density of quarks and gluons in the proton becoming large. The dynamics of the resulting ultra-dense parton system has to differ from that of the dilute system at larger $x$. Indeed, as was observed by Gribov, Levin and Ryskin \cite{Gribov:1984tu} shortly after the BFKL equation was derived, at very high parton density the parton mergers have to eventually compensate for splittings, leading to a slowdown of the PDFs' growth with decreasing $x$. This phenomenon is known as the parton saturation. 

Below we describe how the saturation of gluons leads to unitarization of QCD cross sections, thus allowing us to understand the high-energy asymptotics of strong interactions. The reader interested in a more detailed exposition of the material presented in this Section is referred to \cite{Kovchegov:2012mbw} and references therein.


\subsection{Deep Inelastic Scattering in the quasi-classical approximation}

\begin{figure}
\begin{center}
\includegraphics[width=  \textwidth]{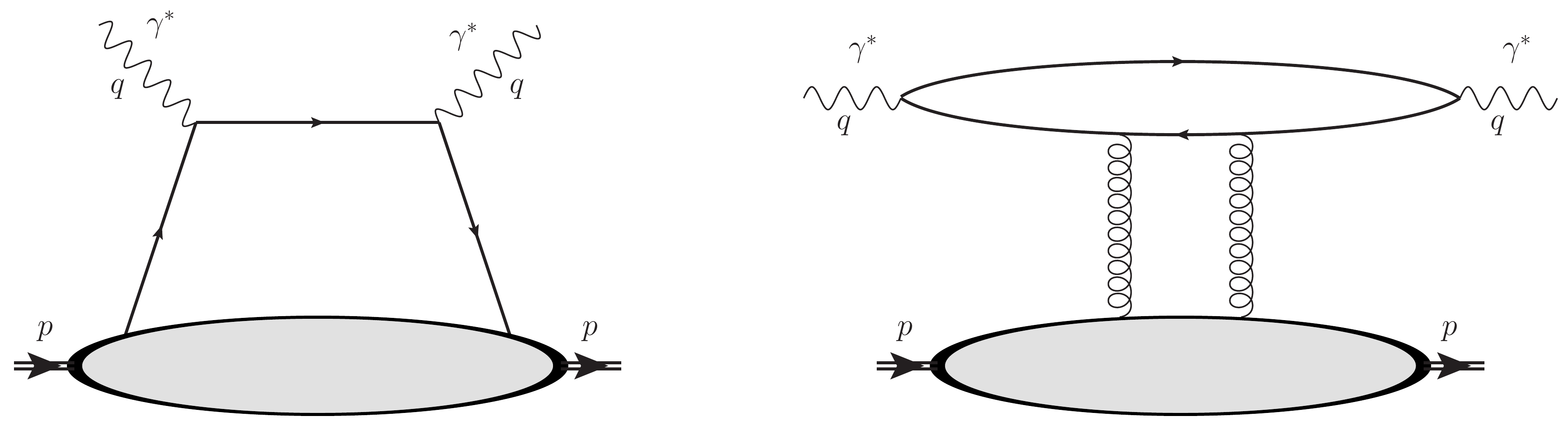} 
\caption{DIS process in the standard ``handbag" diagram approximation (left panel) and due to the exchange of gluons in the $t$-channel (right panel).}
\label{fig:DIS}
\end{center}
\end{figure}

Consider the Deep Inelastic Scattering (DIS) process, where a virtual photon (projectile) scatters on a proton or a nucleus (target). Our usual way of thinking about DIS is in terms of the ``handbag" diagram pictured in the left panel of \fig{fig:DIS}, where the oval at the bottom represents the target proton or nucleus, the wavy line is the virtual photon ($\gamma^*$), and the straight lines are quark propagators. The diagrams in \fig{fig:DIS} contribute to the forward $\gamma^* + p$ scattering amplitude, with the center of mass energy squared $s = (p+q)^2$ of the virtual photon scattering on the target. The leading-order ``handbag" diagram results in the DIS structure functions $F_1$ and $F_2$ expressed in terms of the quark distribution in the target. 

However, if one is interested in DIS at small values of the Bjorken $x$ variable (corresponding to large center-of-mass energy squared $s$ for fixed photon virtuality $Q^2 = - q^2$ with $x \approx Q^2/s$), then a straightforward calculation of the Born-level quark and gluon scattering cross sections demonstrates that $t$-channel exchanges of gluons dominate over the $t$-channel exchanges of quarks. Indeed, if one compares the right panel in \fig{fig:DIS} to the ``handbag" diagram on the left of this figure, one sees that, being a one-loop correction, the right diagrams is suppressed by a power of the strong coupling constant $\as$, while simultaneously being enhanced by a power of $1/x$. That is, for the two diagrams in \fig{fig:DIS} we have
\begin{align}
\frac{\mbox{quark loop diagram}}{\mbox{``handbag" diagram}} \propto \frac{\as}{x}. 
\end{align} 
Clearly, for $x < 0.01$, the $1/x$ enhancement overcomes the $\sim \as$ suppression factor, and the quark loop diagram in \fig{fig:DIS} dominates. The physics behind the quark loop diagram in \fig{fig:DIS}, evaluated at small $x$, is as follows: the virtual photon (produced by the scattering lepton in DIS) fluctuates into a quark--anti-quark pair long before the interaction with the target proton or nucleus. The resulting quark--anti-quark {\sl dipole} goes on to interact with the target. This is the dipole picture of DIS \cite{Gribov:1968gs,Bjorken:1973gc,Bertsch:1981py,Frankfurt:1988nt,Kopeliovich:1981pz,Mueller:1989st,Nikolaev:1990ja}. This interaction is limited to a two-gluon exchange in \fig{fig:DIS}, but can be more complicated in general. Since we are interested in the forward scattering amplitude, long after the interaction the quark--anti-quark pair recombines back into the virtual photon. This interpretation is valid in the rest frame of the target or in the frame where both the virtual photon and the target are relativistic, moving along two different light-cones.  

\begin{figure}
\begin{center}
\includegraphics[width= 0.6 \textwidth]{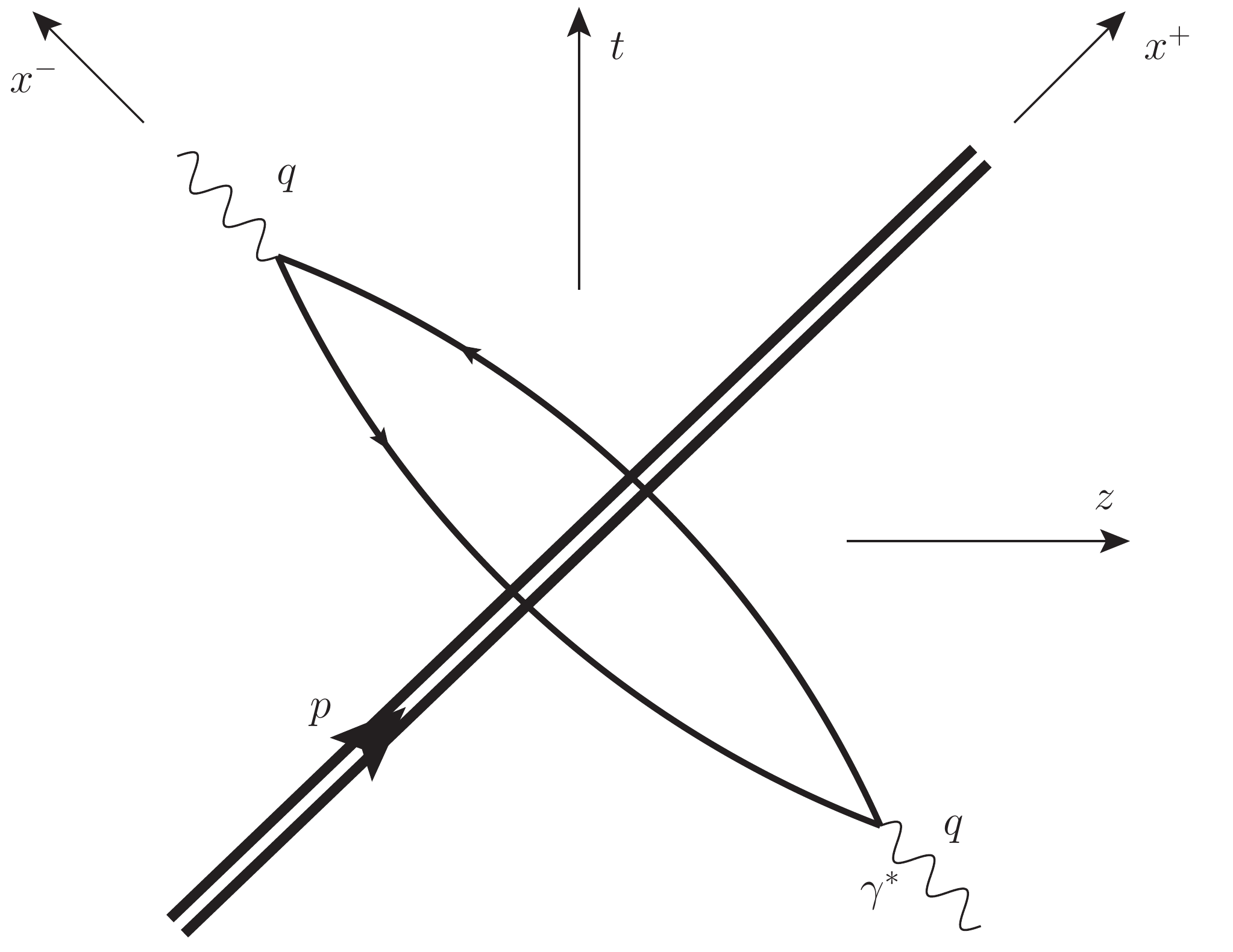} 
\caption{DIS process at small $x$ in the space-time representation.}
\label{fig:DIS_st}
\end{center}
\end{figure}

The space-time picture of the process from the right panel of \fig{fig:DIS} is shown in \fig{fig:DIS_st}. We are working in the frame where the proton and virtual photon momenta are 
\begin{align}
p^\mu \approx (p^+, 0, 0_\perp), \ \ \ q^\mu = \left( - \frac{Q^2}{2 \, q^-} , q^- , 0_\perp \right)
\end{align}
with $p^+$ and $q^-$ large. Our light-cone components are defined by $x^\pm = (x^0 \pm x^3)/\sqrt{2}$ while the transverse vectors are denoted by ${\vec x}_\perp = (x^1, x^2)$ with $x_\perp = |{\vec x}_\perp|$. Figure \ref{fig:DIS_st} illustrates that the lifetime (or, more precisely, $x^-$-extent) of the quark loop is much longer than the width of the Lorentz-contracted proton in the $x^-$ direction. The two-gluon interaction between the quark loop and the proton from \fig{fig:DIS} happens instantaneously in $x^-$ (near $x^- =0$), and does not violate the separation of time scales into the $\gamma^* \to q \bar q$ fluctuation, followed by the interaction and the $q {\bar q} \to \gamma^*$ recombination. Further, one can argue that the high energy interaction of the quark--anti-quark dipole and the proton does not change the transverse positions of the quark and the anti-quark. That is, the interaction is diagonal in the transverse plane \cite{Kopeliovich:1981pz,Levin:1987xr,Mueller:1989st,Nikolaev:1990ja}. We therefore write down the following formula for the DIS cross section at small $x$:
\begin{align}\label{dip_factTL}
  \sigma_{T, \, L}^{\gamma^* A} (x, Q^2) \, = \, \int \frac{d^2
    x_\perp}{4 \, \pi} \, \int\limits_0^1 \, \frac{dz}{z \, (1-z)} \ 
  |\Psi_{T, \, L}^{\gamma^* \rightarrow q {\bar q}} ({\vec x}_\perp,
  z)|^2 \ \sigma_{tot}^{q {\bar q} A} ({\vec x}_\perp, Y),
\end{align}
where subscripts $T, L$ denote the transverse and longitudinal polarizations of the virtual photon. While the target is labeled by the superscript $A$, which, in general, indicates a nucleus, it is also understood that \eq{dip_factTL} applies to DIS on a proton as well. The functions $\Psi_{T, \, L}^{\gamma^* \rightarrow q {\bar q}} ({\vec x}_\perp, z)$ are the light-cone wave functions of the virtual photon fluctuating into the $q \bar q$ pair. They can be calculated using the rules of the light-cone (or light-front) perturbation theory \cite{Lepage:1980fj,Brodsky:1997de}. The wave functions depend on the fraction $z$ of the photon's light-cone momentum $q^-$ carried by the quark in the pair and on the transverse separation ${\vec x}_\perp$ between the quark and the anti-quark. At the leading order one has \cite{Bjorken:1970ah,Nikolaev:1990ja}
\begin{align}\label{Psi2Tx}
  |\Psi_{T}^{\gamma^* \rightarrow q {\bar q}} & ({\vec x}_\perp, z)|^2
  \, = \, 2 \, N_c \, \sum_f \, \frac{\alpha_{EM} \, Z_f^2}{\pi} \, z
  \, (1-z) \notag \\ & \times \, \left\{ a_f^2 \, \left[ K_1 (x_\perp
      \, a_f) \right]^2 \, [z^2 + (1 - z)^2] + m_f^2 \, \left[ K_0
      (x_\perp \, a_f) \right]^2 \right\}, 
\end{align}
and 
\begin{align}\label{Psi2Lx}
  |\Psi_{L}^{\gamma^* \rightarrow q {\bar q}} ({\vec x}_\perp, z)|^2
  \, = \, 2 \, N_c \sum_f \frac{\alpha_{EM} \, Z_f^2}{\pi} \, 4 \, Q^2
  \, z^3 \, (1 - z)^3 \, \left[ K_0 (x_\perp \, a_f) \right]^2,
\end{align}
where 
\begin{align}\label{af}
  a_f^2 \, = \, Q^2 \, z \, (1 - z) + m^2_f.
\end{align}
Here $N_c$ is the number of quark colors, $m_f$ is the mass of quarks with flavor $f$, $Z_f$ is the fractional charge of the quark (in the units of the absolute value of the electron charge), and $\alpha_{EM}$ is the fine-structure constant. 

Employing \eq{dip_factTL} one can construct the DIS structure function via 
\begin{subequations}\label{F12sig}
\begin{align}
  & F_2 (x, Q^2) \, = \, \frac{Q^2}{4 \, \pi^2 \, \alpha_{EM}} \,
  \sigma_{tot}^{\gamma^* A} \, = \, \frac{Q^2}{4 \, \pi^2 \,
    \alpha_{EM}} \, \left[ \sigma_T^{\gamma^* A} + \sigma_L^{\gamma^*
      A} \right], \\
  & 2 \, x \, F_1 (x, Q^2) \, = \, \frac{Q^2}{4 \, \pi^2 \,
    \alpha_{EM}} \ \sigma_T^{\gamma^* A}.
\end{align}
\end{subequations}

The interaction between the dipole and the proton or nucleus is described by the cross section $\sigma_{tot}^{q {\bar q} A} ({\vec x}_\perp, Y)$ in \eq{dip_factTL}, which depends on ${\vec x}_\perp$ and rapidity $Y = \ln (1/x) \approx \ln (s/Q^2)$. All the strong interaction dynamics for the DIS process at small $x$ is contained in this cross section. For further transparency, let us first rewrite this cross section as an integral over the impact parameter ${\vec b}_\perp$, 
\begin{align}\label{s_tot}
\sigma_{tot}^{q {\bar q} A} ({\vec x}_\perp, Y) = 2 \int d^2 b_\perp \, \left[ 1 - \mbox{Re} \, S ({\vec x}_\perp, {\vec b}_\perp, Y) \right]. 
\end{align}
Here $S ({\vec x}_\perp, {\vec b}_\perp, Y)$ is the forward (in the transverse position space) matrix element of the $S$-matrix of the dipole-target scattering, defined by
\begin{align}
S ({\vec x}_\perp, {\vec b}_\perp, Y) = 1 + i \, A ({\vec x}_\perp, {\vec b}_\perp, Y) ,
\end{align}
with 
\begin{align}
A = \frac{M}{2 \, s},
\end{align}
where $M$ is the dipole-target forward scattering amplitude. 

At the leading order in high-energy scattering $A$ is always imaginary, and, hence, $S$ is real. Therefore, we will drop the Re sign in \eq{s_tot}. Defining the imaginary part of $A$ by  $N=\mbox{Im} A$ such that
\begin{align}\label{NvsS}
N ({\vec x}_\perp, {\vec b}_\perp, Y) = 1 -  S ({\vec x}_\perp, {\vec b}_\perp, Y)
\end{align}
we rewrite \eq{s_tot} as
\begin{align}\label{forw_amp}
  \sigma_{tot}^{q {\bar q} A} ({\vec x}_\perp, Y) \, = \, 2 \, \int
  d^2 b \ N ({\vec x}_\perp, {\vec b}_\perp, Y).
\end{align}

At this point let us put our calculation momentarily on hold, and spend a little time discussing unitarity constraints. Unitarity requires that $|S| \le 1$. This leads to $0\le N \le 2$ and 
\begin{align}
\sigma_{tot}^{q {\bar q} A} ({\vec x}_\perp, Y)  \le 4 \, \int d^2 b  = 4 \pi R^2,
\end{align} 
where $R$ is the radius of the (circular) region in the transverse plane where the interaction is strong enough for $S \approx -1$ and $N \approx 2$. This region of strong scattering is known as the {\sl black disk}. In the case of high energy scattering one can argue that, due to the dominance of inelastic processes over the elastic once, $N \le 1$ and 
\begin{align}\label{bdl}
\sigma_{tot}^{q {\bar q} A} ({\vec x}_\perp, Y)  \le 2 \, \int d^2 b  = 2 \pi R^2,
\end{align} 
where now $S \approx 0$ and $N \approx 1$ in the black disk region. \eq{bdl} is known as the black disk limit for the total cross section. Observing that at high energy the radius of the black disk in QCD grows at most logarithmically with energy, $R = R_0 + a \, \ln (s/s_0)$, one arrives at the Froissart-Martin bound on the total cross section \cite{Heisenberg:1952zz,Froissart:1961ux,Martin:1969ina}
\begin{align}\label{FM}
\sigma_{tot}^{q {\bar q} A} ({\vec x}_\perp, Y)  \le 2 \pi a^2 \, \ln^2 \frac{s}{s_0}.
\end{align}
Here $a \sim 1/m_\pi$ with $m_\pi$ the pion mass.

Returning to \eq{forw_amp}, one can show that the two-gluon exchange from the right panel of \fig{fig:DIS} gives
\begin{align}\label{N1nuc}
  N_{LO} ({\vec x}_\perp, {\vec b}_\perp, Y) & \, =  \, \frac{\pi \,
    \as^2 \, C_F}{N_c} \, T({\vec b}_\perp) \, x_\perp^2 \, \ln
  \frac{1}{x_\perp \, \Lambda},
\end{align}
where $C_F = (N_c^2 - 1)/(2 N_c)$ is the fundamental Casimir of SU($N_c$) and $\Lambda$ is some infrared (IR) cutoff. In arriving at \eq{N1nuc} we had to sum over all possible connections of the gluons to the quark and anti-quark line in the dipole. Equation \eqref{N1nuc} is valid for $x_\perp < 1/\Lambda$, that is, the dipole should be perturbatively small. We assumed the target to be a single quark in the proton. The latter could be inside a nucleus. In preparation for the scattering on a nucleus, we have also employed the {\sl nuclear profile function} $T ({\vec
  b}_\perp)$ defined by
\begin{align}\label{npf_def}
  T ({\vec b}_\perp) \, \equiv \, 
  \int\limits_{-\infty}^\infty \, db_3 \, \rho_A ({\vec b}_\perp, b_3)
\end{align}
with $ \rho_A ({\vec b}_\perp, b_3)$ the number density of the nucleons in the nucleus normalized as $\int d^3 b \, \rho_A = A$, where $A$ is the atomic number of the nucleus (the number of protons and neutrons in the nucleus). For the proton target $A=1$. For a large nucleus $T ({\vec b}_\perp) \, \sim \, A^{1/3}$ and the nuclear profile function may become numerically large.  

It appears that we have an immediate problem: as we increase $x_\perp$, the forward amplitude $N_{LO}$ from \eq{N1nuc} grows, potentially exceeding the black-disk limit ($N \le 1$) for some dipole sizes. One may counter this worry by suggesting that dipole sizes should be non-perturbatively large, $x_\perp \gtrsim 1/\Lambda$, for this violation of the black-disk limit to happen. However, for large nuclei with $A \gg 1$ this is not so. In fact, no matter how small the dipole size $x_\perp$ is, we can always increase $A$ until $N_{LO} > 1$ in \eq{N1nuc}. (Indeed this may require unrealistically large nuclei, but this is a theoretical exercise.) Conversely, for a fixed large $A$, as we increase $x_\perp$ we may reach the region where $N_{LO} > 1$ while still being in the perturbative QCD regime of relatively small dipole sizes, $x_\perp < 1/\Lambda$.

\begin{figure}
\begin{center}
\includegraphics[width= 0.78 \textwidth]{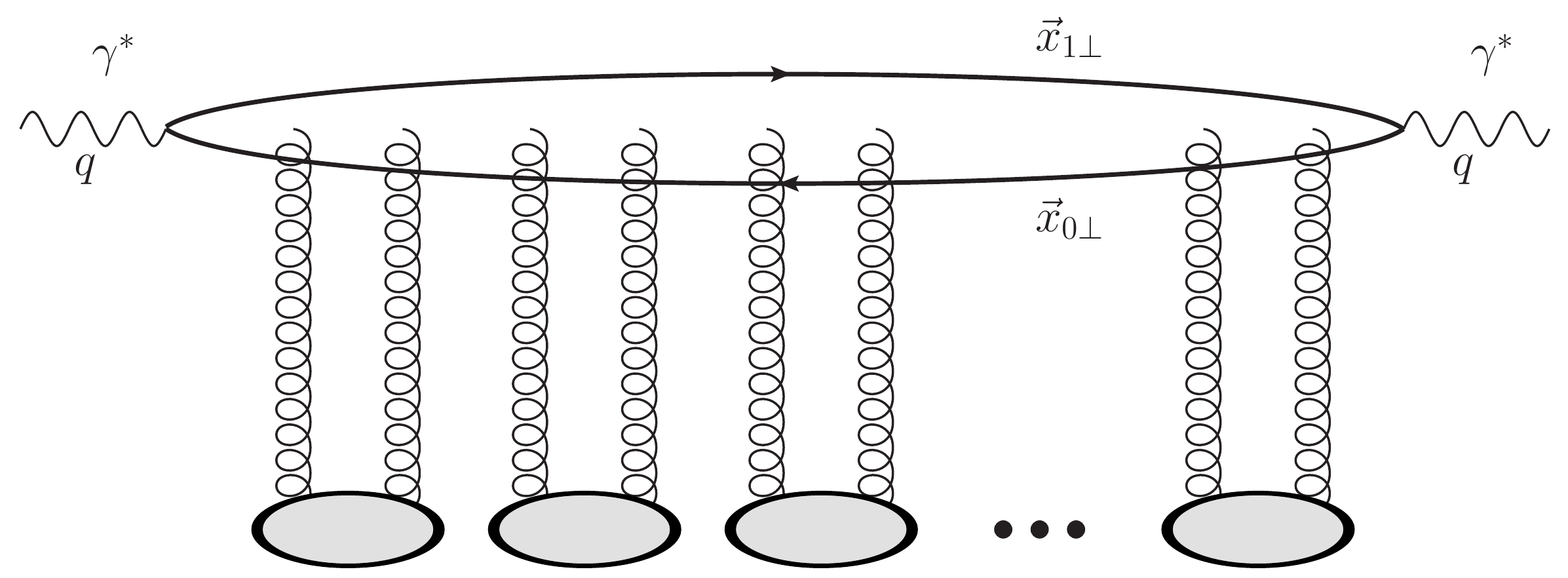} 
\caption{DIS on a nucleus with multiple rescattering on the nucleons (ovals at the bottom). Disconnected gluon lines at the top denote a sum over all possible connections of the gluons to the quark and anti-quark lines.}
\label{fig:GGM}
\end{center}
\end{figure}

In the case of a very large nucleus, our problem above can be resolved in the following way. Large nucleus has many nucleons. When $N_{LO}$ approaches unity, interaction with each nucleon happens with the probability close to one. This means, interactions with multiple nucleons are also becoming likely. Hence, as $N_{LO}$ increases, we need to consider multiple rescatterings of the $q \bar q$ dipole in the nucleus, since those are becoming order-one corrections to the single scattering contained in $N_{LO}$ from \eq{N1nuc}. Diagrammatically we need to sum up the graphs of the type shown in \fig{fig:GGM}, where each oval at the bottom represents a nucleon in the nucleus, and disconnected gluon lines at the top imply summation over all possible connections to the quark and anti-quark lines. 

Interaction with each nucleon is local in $x^-$, such that the gluon exchanges with different nucleons do not get entangled and we also do not have any gluon-gluon interactions. (We are working in the Feynman gauge, $\partial_\mu A^\mu =0$ or in the light-cone gauge of the projectile, $A^- =0$.) This means that we exchange exactly two gluons with each nucleon, as shown in \fig{fig:GGM}. This allows us to define the resummation parameter: two gluons bring in $\as^2$, while there are $\sim A^{1/3}$ nucleons in the nucleus at a given impact parameter. Hence, the resummation parameter of the multiple rescattering calculation is $\as^2 \, A^{1/3}$ \cite{Kovchegov:1997pc}. When $\as^2 \, A^{1/3} \gtrsim 1$, multiple rescatterings are important and have to be re-summed to all orders.  

The result of resumming diagrams in \fig{fig:GGM} is a simple exponentiation of \eq{N1nuc}. It reads \cite{Mueller:1989st}
\begin{align}\label{N_GM2}
  N ({\vec x}_\perp, {\vec b}_\perp, Y =0) \, = \, 1 - \exp \left\{ -
    \frac{\as^2 \, C_F \, \pi}{N_c} \, T({\vec b}_\perp) \, x_\perp^2
    \, \ln \left( \frac{1}{x_\perp \, \Lambda} \right) \right\}.
\end{align}
Defining {\sl the saturation scale} by 
\begin{align}\label{qsmv}
  Q_s^2 ({\vec b}_\perp) \, \equiv \, \frac{4 \, \pi \, \as^2 \,
    C_F}{N_c} \, T({\vec b}_\perp)
\end{align}
we rewrite \eq{N_GM2} as
\begin{align}\label{glaN2}
  N({\vec x}_\perp, {\vec b}_\perp , Y=0) \, = \, 1 - \exp \left\{ -
    \frac{1}{4} \, x_\perp^2 \, Q_s^2 ({\vec b}_\perp) \, \ln \frac{1}{x_\perp \,
      \Lambda} \, \right\}.
\end{align}
This is the Gribov--Glauber--Mueller (GGM) \cite{Mueller:1989st} formula for the dipole amplitude $N$.

\begin{figure}
\begin{center}
\includegraphics[width= 0.65 \textwidth]{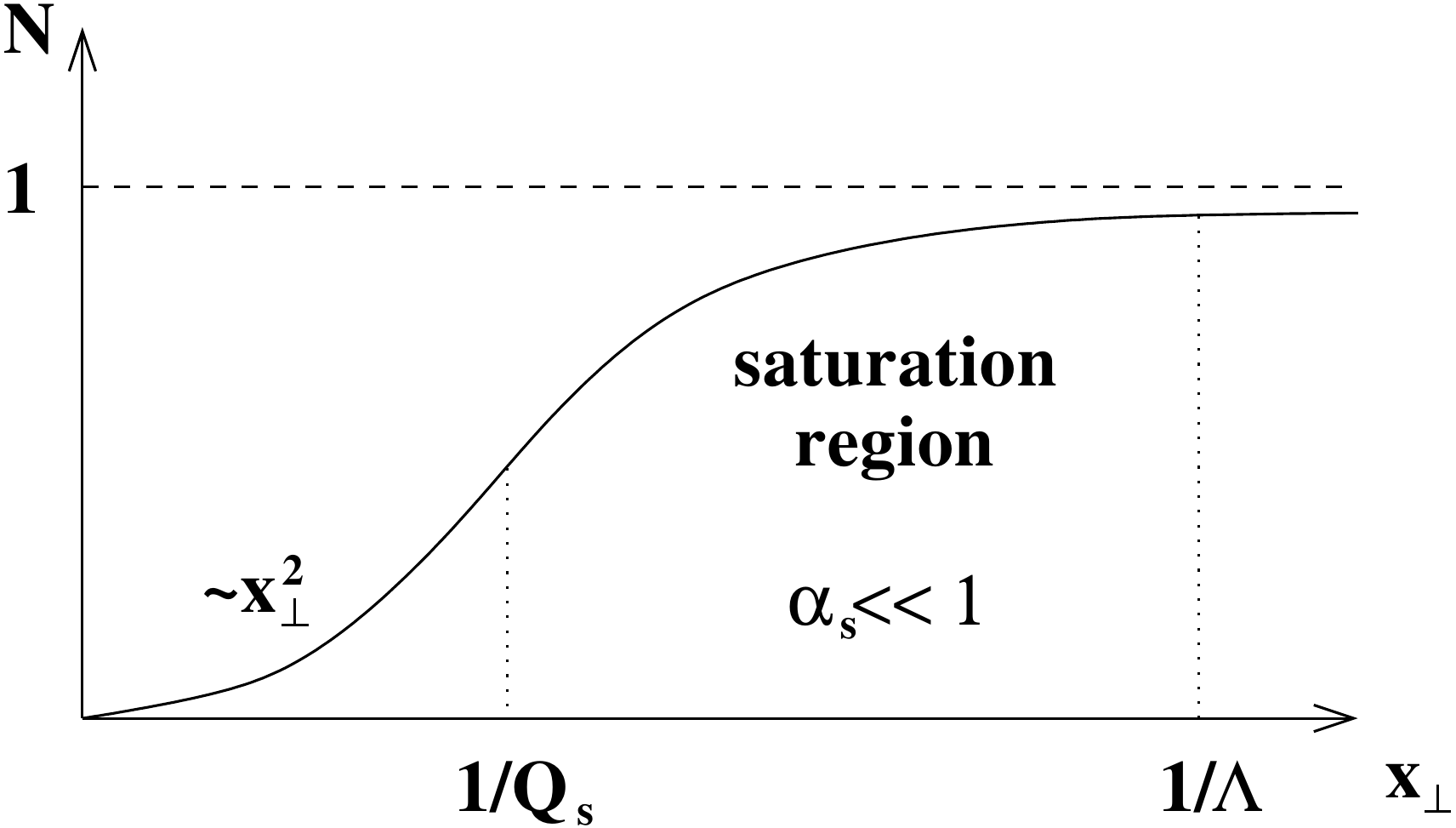} 
\caption{The forward scattering amplitude given by \eq{glaN2} plotted as a function of the dipole transverse size $x_\perp$.}
\label{fig:saturation}
\end{center}
\end{figure}

The GGM formula is sketched in \fig{fig:saturation} as a function of $x_\perp$. We see that, just like for $N_{LO}$ from \eq{N1nuc}, the amplitude $N$ from \eq{glaN2} approaches zero as $x_\perp \to 0$. Indeed, this makes perfect physical sense: when the quark and the anti-quark are close to each other, their colors cancel and their interaction with the target  tends to zero. This phenomenon is known as color transparency \cite{Kopeliovich:1981pz,Nikolaev:1990ja,Heiselberg:1991is,Frankfurt:1993it}.

Remarkably, when $x_\perp$ increases, the amplitude in \eq{glaN2} and \fig{fig:saturation} never exceeds one, that is, one always has $N<1$ and the black disk limit is not violated (in the perturbative region, $x_\perp < 1/\Lambda$). This behavior is due to the exponentiation in \eq{glaN2}, that is, due to multiple rescatterings. We see that inclusion of multiple rescatterings indeed resolved the problem of the black-disk-limit violation and of unitarity violation by the leading-order expression \eqref{N1nuc}. 

Since $T({\vec b}_\perp) \sim A^{1/3}$ we have the following scaling:
\begin{align}\label{QsA}
  Q_s^2 \, \sim \, A^{1/3}. 
\end{align}
Therefore, for large enough nucleus, one has $Q_s^2 \gg \Lambda^2$, and the saturation scale is perturbatively large. Since the transition from the region where $N$ is rising with increasing $x_\perp$ to the region where the growth is gradually tamed happens at $x_\perp \approx 1/Q_s  \ll 1/\Lambda$, the unitarization takes place in the perturbative region where $\as \ll 1$ still. The region $1/Q_s \lesssim x_\perp \ll 1/\Lambda$ is called {\sl the saturation region}. There the strong coupling constant $\as$ is small, but, due to the large size of the nucleus, the interactions are strong and the resulting dipole--nucleus cross section approaches the black disk limit of $N=1$ (without exceeding it). We conclude that the saturation region appears to be weakly-coupled, but strongly-interacting. 

The GGM approach resumming the parameter $\as^2 \, A^{1/3}$ is equivalent to the quasi-classical description of the small-$x$ nuclear wave function in the framework of the McLerran--Venugopalan (MV) model  \cite{McLerran:1993ni,McLerran:1993ka,McLerran:1994vd}. Gluon fields in the MV model are classical at the leading order in the resummation of powers of $\as^2 \, A^{1/3}$, and have to be found by solving the classical Yang--Mills equations. Further details on this very important model for saturation physics can be found in the reviews \cite{Iancu:2003xm,Weigert:2005us,Jalilian-Marian:2005jf,Gelis:2010nm,Albacete:2014fwa} and in the book \cite{Kovchegov:2012mbw}.

Last but not least, let us formulate our results in terms of operators. The propagator for a high-energy quark (or anti-quark) can be written as (proportional to) an eikonal Wilson line along the corresponding light-cone trajectory. Since the quark and anti-quark are generated in our small-$x$ scattering process long before the interaction with the target, and are absorbed back into the virtual photon long after that interaction, we can think of their (close to) light-cone trajectories as being infinite in $x^-$. Define a light-cone Wilson line for quarks by
\begin{align}\label{Vline}
V_{{\vec x}_\perp} [b^-, a^-] = \mbox{P} \exp \left\{ i g \int\limits_{a^-}^{b^-} d x^- \, A^+ (x^+ =0, x^-, {\vec x}_\perp ) \right\} 
\end{align}
with $A^\mu = t^a \, A^{a \mu}$ and $t^a$ the generators of SU($N_c$) in the fundamental representation. Infinite Wilson line is denoted by 
\begin{align}\label{Vline_inf}
V_{{\vec x}_\perp}  \equiv V_{{\vec x}_\perp} [+\infty, -\infty].  
\end{align}
With the help of these operators, we can write down the expectation value for the $S$-matrix of the dipole--target scattering as
\begin{align}\label{Sdef}
S ({\vec x}_{1 \perp}, {\vec x}_{0 \perp}, Y) = \left\langle \frac{1}{N_c} \, \mbox{tr} \left[ V_{{\vec x}_{1 \perp}} \, V^\dagger_{{\vec x}_{0 \perp}} \right]  \right\rangle . 
\end{align}
We have switched notation in \eq{Sdef}: instead of using the dipole separation ${\vec x}_\perp$ and the impact parameter ${\vec b}_\perp$, we now are using the transverse positions of the quark and the anti-quark, 
\begin{align}
 {\vec x}_{1 \perp} = {\vec b}_{\perp} + \thalf \, {\vec x}_{\perp}, \ \ \ {\vec x}_{0 \perp} = {\vec b}_{\perp} - \thalf \, {\vec x}_{\perp},
\end{align}
in the argument of $S$, as shown in \fig{fig:GGM}. 

The angle brackets in \eq{Sdef} indicate averaging in the target proton or nucleus state (see e.g. \cite{Kovchegov:1996ty}). They can be related to the standard expectation value in the proton (nucleus) state by 
\begin{align}\label{matrix_el3}
\left\langle \hat{\cal O} (b, r) \right\rangle = \frac{1}{2 p^+} \, \int \frac{d^2 \Delta_\perp \, d \Delta^+}{(2 \pi)^3} \, e^{i b \cdot \Delta} \, \bra{p+ \frac{\Delta}{2}} \hat{\cal O} (0,r) \ket{p - \frac{\Delta}{2}} 
\end{align}
for an arbitrary operator $\hat{\cal O} (b, r)$ \cite{Kovchegov:2019rrz}. Here $b \cdot \Delta = b^- \Delta^+ - {\vec b}_\perp \cdot {\vec \Delta}_\perp$. 

With the help of \eq{Sdef} we re-write the dipole scattering amplitude as
\begin{align}\label{Ndef}
N ({\vec x}_{1 \perp}, {\vec x}_{0 \perp}, Y) = 1 - \left\langle \frac{1}{N_c} \, \mbox{tr} \left[ V_{{\vec x}_{1 \perp}} \, V^\dagger_{{\vec x}_{0 \perp}} \right]  \right\rangle . 
\end{align}
Equation \eqref{glaN2} evaluates the correlation function in \eq{Ndef} in the quasi-classical GGM/MV approximation. 

In the operator language, formula \eqref{dip_factTL} can be interpreted as a statement of factorization between the light-cone wave functions squared (the so-called impact factor) and the correlation function of two infinite light-cone Wilson lines. The latter contains the bulk of the strong interactions dynamics. It is interesting to note that this factorization, while naturally appearing in this quasi-classical calculation due to the separation between the $\gamma^* \to q \bar q$ splitting time scale and the interaction time scale, will remain valid once the small-$x$ evolution corrections are included.  

The operator definitions \eqref{Sdef} and \eqref{Ndef} are strictly-speaking not gauge-invariant: to make them gauge-invariant one needs to insert into the trace a pair of transverse-pointing gauge links connecting the two light-cone Wilson lines at the infinities $x^- = \pm \infty$. However, in the $A^- = 0 $ and $\partial_\mu A^\mu =0$ gauges that we employ here such gauge links do not contribute: thus, we will follow the standard notation and will not show these links explicitly. (The gauge links at infinity become very important in $A^+ =0$ gauge.) 

The operator formalism we have just introduced may not seem particularly useful in the quasi-classical limit considered in this Section. It is, nevertheless, a very powerful method, whose benefits we will demonstrate later in this Chapter.


\subsection{Nonlinear small-$x$ BK evolution equation at large-$N_c$ and Mueller's dipole model}

The GGM formula \eqref{glaN2} derived above describes how saturation effects prevent the cross section from violating the black disk limit. However it suffers from one significant shortcoming: the amplitude $N$ given by the GGM formula is independent of the center-of-mass energy, or, equivalently, of rapidity $Y$. Indeed the saturation scale \eqref{qsmv} is energy-independent, and so is the amplitude $N$ in \eq{glaN2}. This is why we put $Y=0$ in the argument of $N$ in \eq{glaN2}: we will shortly see that this implies that \eq{glaN2} will serve as the initial condition for the evolution equation we are about to derive. This energy independence of the GGM formula \eqref{glaN2} contradicts experimental data: for instance, the measured $F_2$ structure function is indeed a function of $x$, which grows rapidly with decreasing $x$. Theoretically, it would be important to find radiative corrections to \eq{glaN2}: above they were neglected as being suppressed by an additional power of $\as$, and not enhanced by $A^{1/3}$. However, at small $x$ there exists another large parameter, $\ln (1/x)$, and some of the radiative corrections to the GGM formula come in with a factor of $\as \, \ln (1/x)$. That is, the suppression introduced by $\as \ll 1$ can be compensated by the large logarithm $\ln (1/x) \gg 1$, such that the resulting parameter is order one, $\as \, \ln (1/x) \sim 1$, and the powers of this parameter need to be summed up to all orders.  
 
\begin{figure}
\begin{center}
\includegraphics[width= 0.75 \textwidth]{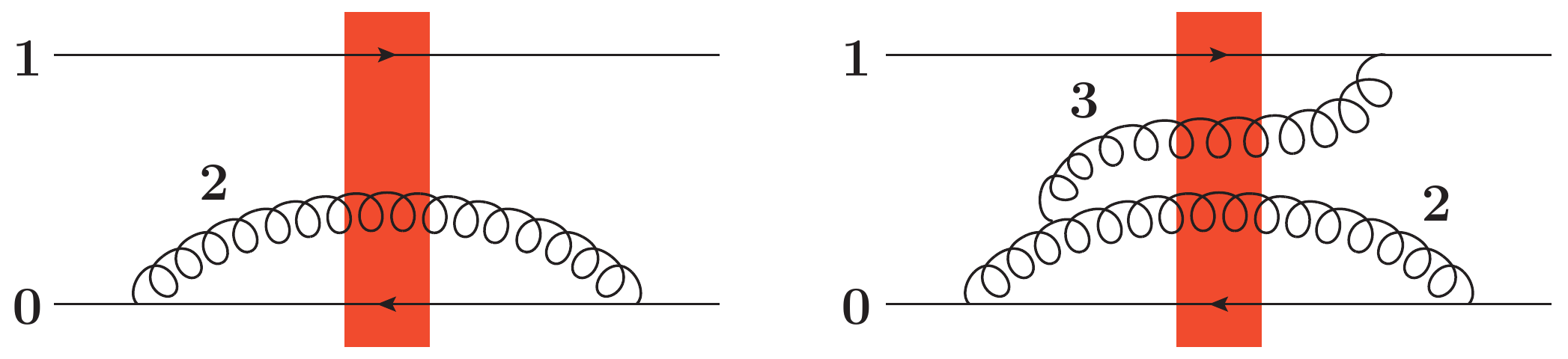} 
\caption{One and two-gluon emissions by the $q \bar q$ dipole scattering on a target shock wave.}
\label{fig:12gluon}
\end{center}
\end{figure}

\begin{figure}
\begin{center}
\includegraphics[width= 0.8 \textwidth]{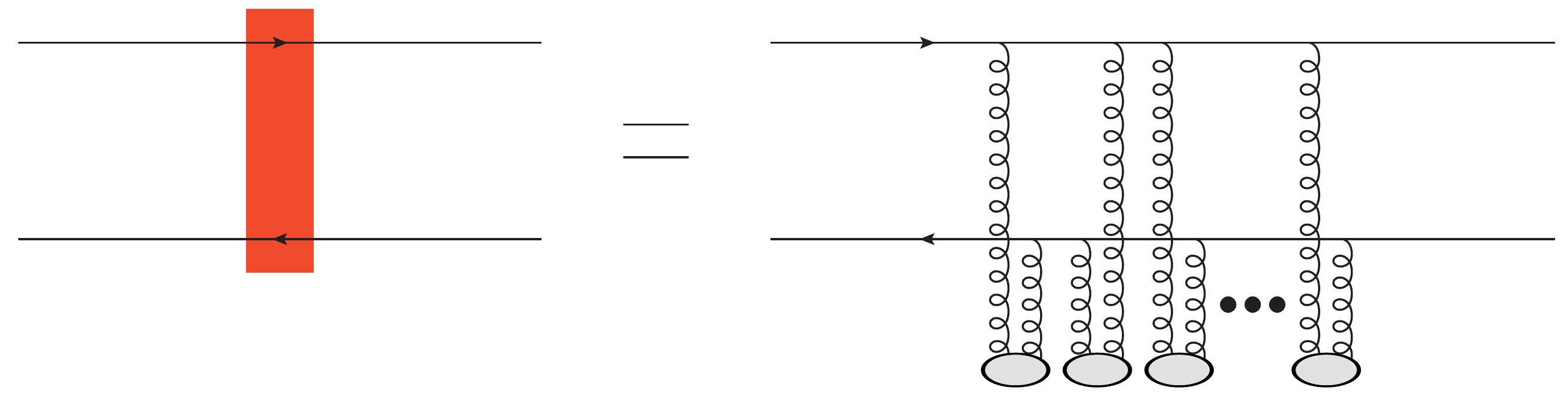} 
\caption{The abbreviated notation for a shock wave, with the red rectangle representing all interactions with the target due to GGM gluons exchanges.}
\label{fig:shock_wave}
\end{center}
\end{figure}

One can show that radiative corrections bringing in powers of $\as \, \ln (1/x)$ in the $A^- =0$ gauge are generated by the long lived gluons emitted (and absorbed) by the projectile dipole (moving in the $x^-$ direction, as per \fig{fig:DIS_st}). This is illustrated in \fig{fig:12gluon}. The $x^-$-lifetime of those gluons should be much longer than the interaction time between the projectile and the target via the GGM gluons. Therefore, we can combine the interaction with the GGM gluons into an almost instantaneous interaction with the ``shock wave", as depicted in by the red rectangle in \fig{fig:12gluon}. The meaning of the red rectangle is clarified in \fig{fig:shock_wave}: at this point it represents all GGM-like interaction of the projectile with the target. If the projectile is not just a $q \bar q$ pair, and if it includes long-lived gluons as in \fig{fig:12gluon}, those gluons can also interact with the target by exchanging instantaneous (GGM) gluons. 

Each long-lived gluon in \fig{fig:12gluon} brings in a power of $\as \, \ln (1/x)$, with the logarithm of $1/x$ arising from the phase space integral as we will see shortly. One can show that the gluons emitted or absorbed when the dipole is inside the shock wave do not generate $\ln (1/x)$: hence we need to consider only the gluons emitted and absorbed when the dipole is outside the shock wave. From \fig{fig:12gluon} we conclude that to resum all powers of $\as \, \ln (1/x)$ we need to sum up an infinite cascade of these long-lived gluons scattering on the shock wave target. 

\begin{figure}
\begin{center}
\includegraphics[width= 0.75 \textwidth]{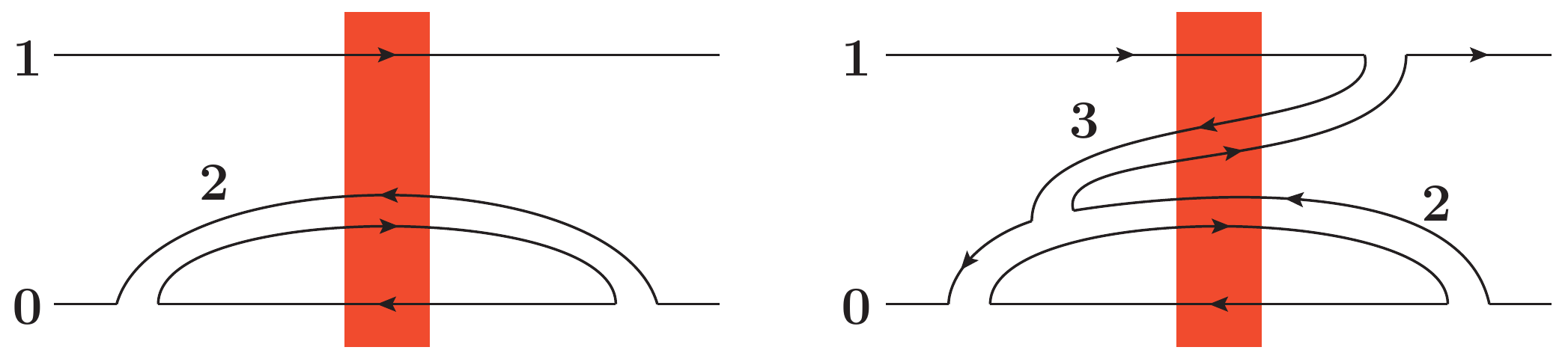} 
\caption{One and two-gluon emissions by the $q \bar q$ dipole scattering on a target shock wave in the large-$N_c$ limit.}
\label{fig:12gluon_LargeNc}
\end{center}
\end{figure}

Resummation of a gluon cascade is not easy in general. To simplify it, we will employ the 't Hooft's large-$N_c$ limit \cite{tHooft:1974pnl} following the prescription of the Mueller's dipole model \cite{Mueller:1993rr,Mueller:1994jq,Mueller:1994gb} (see also \cite{Nikolaev:1993ke}). In the large-$N_c$ limit only the planar diagrams survive. This means that at large $N_c$ the diagrams in \fig{fig:12gluon} can be redrawn as shown in \fig{fig:12gluon_LargeNc}, where the double lines represent the gluons. (At large $N_c$ each gluon can be represented as a quark and anti-quark pair in the $N_c^2 -1$ adjoint representation of SU($N_c$).)

\begin{figure}
\begin{center}
\includegraphics[width= 0.9 \textwidth]{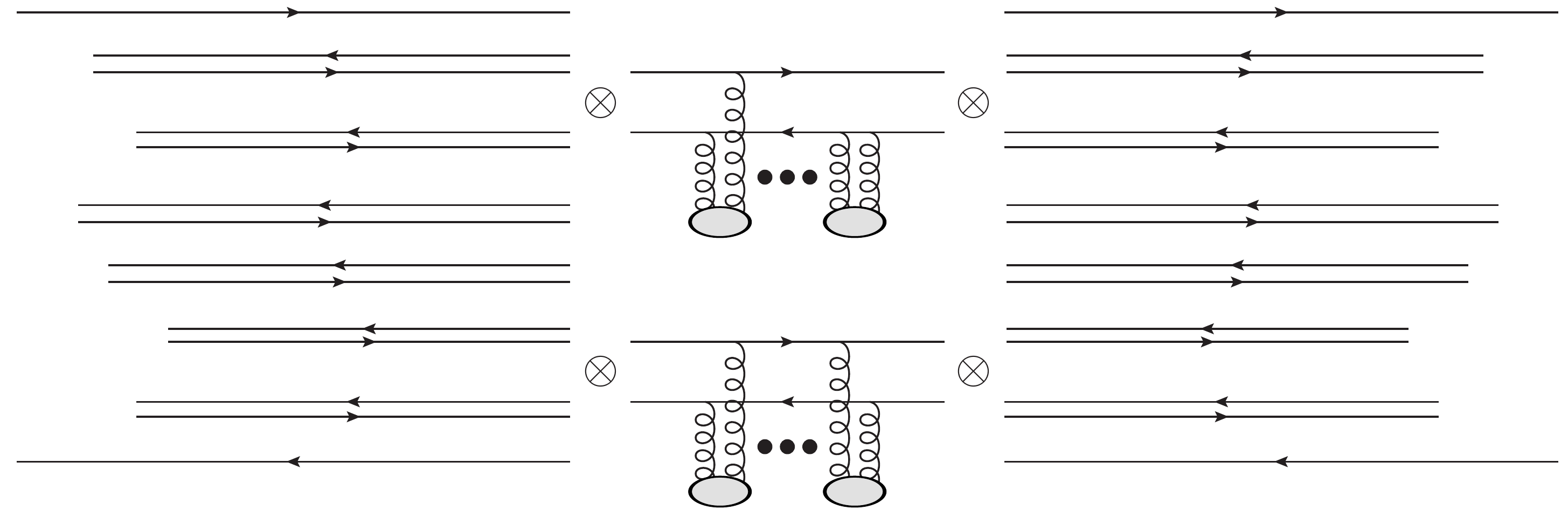} 
\caption{The color dipole cascade, with each dipole interacting independently with the target nucleus.}
\label{fig:cascade}
\end{center}
\end{figure}

Let us start with the left panel in \fig{fig:12gluon_LargeNc}. There, the emission of gluon 2 in the large-$N_c$ limit splits the original color dipole (made out of the quark 1 and the anti-quark 0) into two color dipoles, one made out of the quark 1 and the anti-quark part of the gluon line 2, and the other one made out of the quark part of the gluon line 2 and the anti-quark 0. It is important to point out that, in order to generate logarithms of $x$, the emitted gluon 2 has the minus component of its momentum much smaller than the minus momenta of the quark and anti-quark in the original dipole, $k_2^- \ll k_1^-, k_0^-$. An essential consequence of this ordering is that the transverse position of the anti-quark 0 (in the left panel of \fig{fig:12gluon_LargeNc}) does not change after emitting and absorbing the gluon 2. The same applies to the quark 1 and the gluon 2 in the right panel of \fig{fig:12gluon_LargeNc}: their transverse positions do not change after the gluon 3 emission and absorption due to the $k_3^- \ll k_2^- \ll k_1^-, k_0^-$ ordering. Similar to the left panel, in the right panel of \fig{fig:12gluon_LargeNc} we observe that the original color dipole 10 is split into three color dipoles 13, 32, and 20, by the time the projectile system crosses the shock wave. We see that, at large $N_c$, our gluon cascade turns into a cascade of color dipoles. This is the essence of Mueller's dipole model. Moreover, note that the dipoles do not interact with each other: such interactions would be non-planar, and hence $N_c$-suppressed. We conclude that we need to sum up a cascade of dipoles, in which each dipole may produce other (``daughter") dipoles, but the dipoles do not interact with each other otherwise. In the end, each dipole interacts with the shock wave independently via the GGM multiple rescatterings. This physical picture is illustrated in \fig{fig:cascade}, where the disconnected gluon (double straight) lines represent the sum over all possible (planar) ways each gluon can be emitted by the gluons which were emitted earlier in the cascade. The $\otimes$ signs in \fig{fig:cascade} denote the convolution between the GGM formula \eqref{glaN2} and the dipole cascade over the transverse positions of the quark and anti-quark in the dipoles. (In addition to the gluons crossing the shock wave, as shown in \fig{fig:12gluon_LargeNc}, there may also be virtual gluons, which do not cross the shock wave: those are both emitted and absorbed either to the left or to the right of the shock wave. While these gluons need to be included as well, and will be included shortly, we have not explicitly discussed them here, since they do not modify the qualitative picture of the dipole cascade we have described.)

\begin{figure}
\begin{center}
\includegraphics[width= \textwidth]{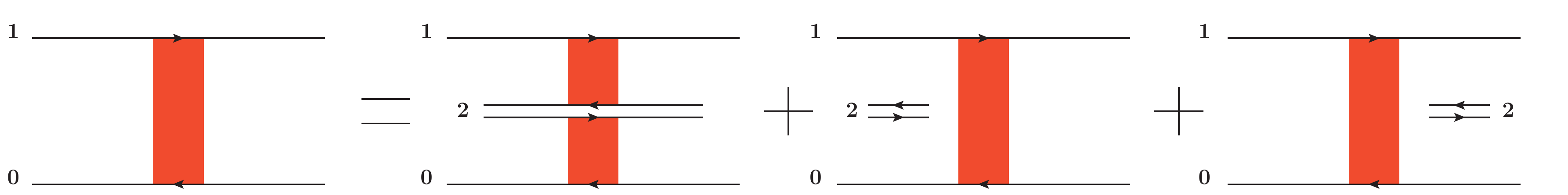} 
\caption{Diagrammatic illustration of the dipole evolution at small $x$ in the large-$N_c$ limit.}
\label{fig:BK}
\end{center}
\end{figure}

We are now ready to resum the cascade. Employing the above observations that the dipoles do not interact with each other and that they also interact with the shock wave target independently, we arrive at the evolution equation demonstrated diagrammatically in \fig{fig:BK}. There we describe one step of small-$x$ evolution for a color dipole. Disconnected gluon (double) lines again denote all possible ways of emitting or absorbing the gluon by the quark and the anti-quark lines in the dipole. The last two diagrams on the right of \fig{fig:BK} denote the virtual gluon correction we have just mentioned above. The meaning of the shock wave has changed as compared to \fig{fig:shock_wave}: now it includes not only the GGM gluon exchanges, but also the subsequent gluon emissions in the cascade. Indeed, as we pointed out before, the minus momenta of the gluons in the cascade are ordered, 
\begin{align}\label{long_ord}
k_0^-, k_1^- \gg k_2^- \gg k_3^- \gg \ldots ,
\end{align}
while the transverse momenta are comparable to each other, 
\begin{align}
k_{0 \perp}, k_{1 \perp} \sim k_{2 \perp} \sim k_{3 \perp} \sim \ldots .
\end{align}
These conditions lead to the ordering of the $x^-$-lifetimes of the cascade gluons, 
\begin{align}\label{lifetime_ord}
\frac{2 k_2^-}{k_{2 \perp}^2} \gg \frac{2 k_3^-}{k_{3 \perp}^2} \gg \ldots .
\end{align}
Since the softer (smaller-$k^-$) gluons have much shorter lifetimes, they can be considered as a part of the shock wave from the standpoint of the harder, longer-living gluons. 

In \fig{fig:BK} we have also modified the shock wave notation to underline the fact that each dipole interacts independently with the shock wave by drawing the shock wave only inside each dipole in the first diagram on the right-hand side. The physical picture of the evolution in \fig{fig:BK} is then rather transparent: in one step of small-$x$ evolution at leading order in $\as$ only one gluon can be emitted. The gluon can be ``real", splitting the original 10 dipole into two dipoles, 12 and 20 by the time the system crosses the shock wave, as shown in the first diagram on the right of \fig{fig:BK}. Alternatively, the gluon can be virtual, and be both emitted and absorbed either to the left or to the right of the shock wave, as shown in the last two diagrams in \fig{fig:BK} respectively. Virtual gluons leave the original dipole 10 to interact with the target shock wave.

An explicit calculation of the diagrams pictured in \fig{fig:BK} leads to the following evolution equation for the dipole $S$-matrix:
\begin{align}\label{Sevol}
  & \partial_Y S ({\vec x}_{1 \perp}, {\vec x}_{0
    \perp}, Y ) \, = \, \frac{\as \, N_c}{2 \, \pi^2} \int d^2 x_2 \,
  \frac{x^2_{10}}{x^2_{20}\,x^2_{21}} \notag \\ & \times \left[ S ({\vec x}_{1 \perp}, {\vec x}_{2
    \perp}, Y )
    S ({\vec x}_{2 \perp}, {\vec x}_{0
    \perp}, Y ) - S ({\vec x}_{1 \perp}, {\vec x}_{0
    \perp}, Y ) \right],
\end{align}
where $\partial_Y \equiv \partial/\partial Y$ and $x_{ij} = |{\vec x}_{i \perp} -  {\vec x}_{j
    \perp}|$ are the transverse sizes of the dipoles. Equation \eqref{Sevol} is the Balitsky--Kovchegov (BK) equation \cite{Balitsky:1995ub,Balitsky:1998ya,Kovchegov:1999yj,Kovchegov:1999ua} for the dipole $S$-matrix. We are using the notation where the arguments of $S$ are the positions of the quark and anti-quark in the dipole, along with $Y = \ln (1/x)$. 

Employing \eq{NvsS}, the BK equation \eqref{Sevol} can be re-written in a somewhat more conventional form as
\begin{align}\label{eqN2}
  &  \partial_Y N ({\vec x}_{1 \perp}, {\vec x}_{0
    \perp}, Y ) \, = \, \frac{\as \, N_c}{2 \, \pi^2} \, \int \, d^2
  x_2 \, \frac{x^2_{10}}{x^2_{20}\,x^2_{21}} \, \bigg[ N ( {\vec
    x}_{1 \perp}, {\vec x}_{2 \perp}, Y ) + N ( {\vec
    x}_{2 \perp}, {\vec x}_{0 \perp} , Y ) \notag \\ & - N ({\vec x}_{1 \perp},
  {\vec x}_{0 \perp}, Y ) - N ( {\vec x}_{1 \perp}, {\vec x}_{2
    \perp}, Y ) \, N ( {\vec x}_{2 \perp}, {\vec x}_{0 \perp}, Y
  ) \bigg].
\end{align}
As we mentioned above, the initial condition for the BK evolution \eqref{eqN2} at $Y=0$ can be given by the GGM formula \eqref{glaN2}. In this case, the solution of the BK equation would resum the powers of $\as N_c Y$ due to the evolution \eqref{eqN2} along with the powers of $\as^2 \, A^{1/3}$ due to the GGM initial condition \eqref{glaN2}. 

We now have the following formalism to calculate DIS structure functions of proton and nuclei at small $x$. One first has to solve the BK equation \eqref{eqN2} using some initial conditions at $Y=0$, preferably the GGM formula \eqref{glaN2}. Then, using the resulting dipole amplitude $N$ in Eqs.~\eqref{forw_amp}, \eqref{dip_factTL}, and \eqref{F12sig}, while also employing Eqs.~\eqref{Psi2Tx} and \eqref{Psi2Lx} one would arrive at predictions for the DIS structure functions $F_2$ and $F_1$. Non-perturbative QCD physics may enter through the initial conditions for the BK evolution. (Though they are suppressed when the saturation scale is sufficiently large.) In addition, the contribution of large impact parameters $b$ in the \eq{forw_amp} may suffer from large non-perturbative corrections \cite{Kovner:2001bh}. However, for a sufficiently large nuclear target such peripheral contributions are suppressed by a power of $A^{1/3}$ and can be neglected. 

If we neglect the quadratic in $N$ term in \eq{eqN2} retaining only the linear terms on its right-hand side (for instance, in the $N \ll 1$ approximation), we would obtain the BFKL equation \cite{Kuraev:1977fs,Balitsky:1978ic} in the dipole form. However, for larger values of the dipole scattering amplitude $N$, the quadratic term in \eq{eqN2} becomes important. Hence, the BK evolution generalizes the BFKL equation to the case of the scattering amplitude approaching the black-disk limit $N \lesssim 1$. The BK equation has the same structure (BFKL equation minus a quadratic term) as the Gribov-Levin-Ryskin (GLR) equation \cite{Gribov:1984tu}, which laid the foundation for the saturation physics. The double logarithmic limit of GLR equation arising at large values of $Q^2$ was independently studied in \cite{Mueller:1985wy}, leading to the GLR-MQ equation.The GLR and GLR-MQ equations were derived by summing the so-called BFKL pomeron ``fan diagrams", with the quadratic term considered to be only the first saturation correction: the BK equation above shows that the quadratic term accounts for all saturation effects at large $N_c$ and in the leading logarithmic approximation (LLA), in which we resum powers of $\as \, \ln (1/x)$. 

An important conclusion we derive from the above derivation of the BK equation is that color dipoles are a useful degree of freedom in high energy QCD at large $N_c$. Instead of gluon degrees of freedom, one seems to be better off using dipoles. Below we discuss how this conclusion can be generalized to light-cone Wilson lines.  

While Lev Lipatov was not actively involved in the saturation research, he, together with collaborators, re-derived the BK equation in \cite{Bartels:2004ef}, establishing an important connection between the $s$-channel evolution involving the shock wave that we presented here and the traditional $t$-channel formalism used in the original derivation of the BFKL equation. 


\subsection{Light cone Wilson line operators as new degrees of freedom; JIMWLK Equation}

\begin{figure}
\begin{center}
\includegraphics[width= \textwidth]{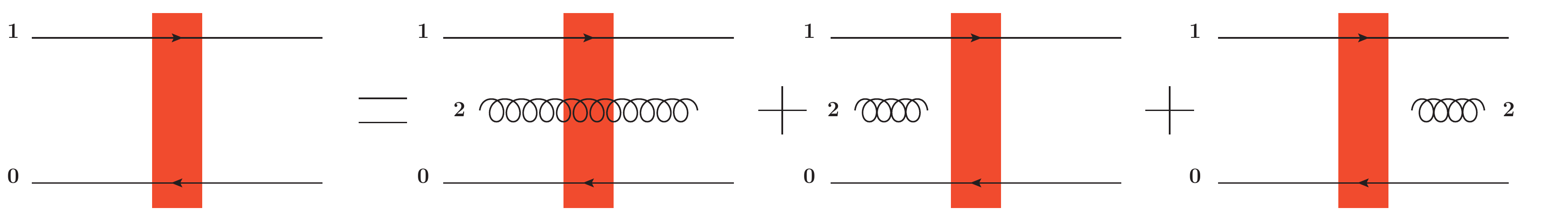} 
\caption{Diagrammatic illustration of one step of the dipole evolution at small $x$ beyond the large-$N_c$ limit.}
\label{fig:Balitsky}
\end{center}
\end{figure}

Let us revisit the above gluon/dipole cascade resummation and try to avoid taking the large $N_c$ limit. Physically very little would change in the above picture. Indeed, instead of a dipole cascade we would have a gluon cascade. However, just like at large $N_c$, the harder gluons would be longer-lived than the softer gluons. Hence the latter can be absorbed in the shock wave. While dipoles are no longer the right degrees of freedom, one can take a color dipole as a starting point and imagine one step of evolution for the dipole in the shock wave background without taking the large-$N_c$ limit. This is depicted in \fig{fig:Balitsky} which looks somewhat similar to \fig{fig:BK}, except the gluon line now is not replaced by the quark an anti-quark lines. Again, disconnected gluon lines imply all possible connections of the gluon 2 to the quark 1 and the anti-quark 0. 

Another similarity between the large-$N_c$ and any-$N_c$ evolution is that the emission of softer gluons does not change the transverse positions of the harder gluons or quarks. Hence, the interaction of quarks and gluons with the shock wave can be described by Wilson lines. Define the adjoint light-cone Wilson line by 
\begin{align}\label{Uline}
U_{{\vec x}_\perp} [b^-, a^-] = \mbox{P} \exp \left\{ i g \int\limits_{a^-}^{b^-} d x^- \, {\cal A}^+ (x^+ =0, x^-, {\vec x}_\perp ) \right\} 
\end{align}
with ${\cal A}^\mu = T^a \, A^{a \mu}$ and $T^a$ the generators of SU($N_c$) in the adjoint representation. Similar to the case of the fundamental Wilson lines \eqref{Vline_inf}, infinite adjoint light-cone Wilson line is denoted by 
\begin{align}\label{Uline_inf}
U_{{\vec x}_\perp}  \equiv U_{{\vec x}_\perp} [+\infty, -\infty].  
\end{align}

In terms of Wilson lines, one step of a dipole evolution depicted in \fig{fig:Balitsky} leads to the following equation:
\begin{align}\label{Bal1}
\partial_Y & \left\langle \frac{1}{N_c} \, \mbox{tr} \left[ V_{{\vec x}_{1 \perp}} \, V^\dagger_{{\vec x}_{0 \perp}} \right]  \right\rangle_Y  = \frac{\as}{\pi^2} \int \, d^2
  x_2 \, \frac{x^2_{10}}{x^2_{20}\,x^2_{21}} \\ & \times \, \left[ \left\langle \frac{1}{N_c} \, \mbox{tr} \left[ t^b \,  V_{{\vec x}_{1 \perp}} \, t^a \, V^\dagger_{{\vec x}_{0 \perp}} \right] \, U_{{\vec x}_{2 \perp}}^{ba} \right\rangle_Y - \left\langle \frac{C_F}{N_c} \, \mbox{tr} \left[ V_{{\vec x}_{1 \perp}} \, V^\dagger_{{\vec x}_{0 \perp}} \right]  \right\rangle \right]_Y . \notag
\end{align}
Here we have explicitly shown the indices $a, b$ of the adjoint Wilson line accounting for the gluon 2 interaction with the shock wave. The rapidity dependence of the Wilson line correlators can be introduced by tilting the Wilson lines contours away from the light cone. It is denoted by the subscript $Y$ on the angle brackets.   

Equation \eqref{Bal1} captures the difficulty one has when trying to generalize the BK evolution to all $N_c$: the first correlator on its right-hand side is not the same as the dipole from its left-hand side (cf. \eq{Sdef}). It is a correlator describing the interaction of a quark, anti-quark and a gluon with the shock wave, as depicted in the first diagram on the right of \fig{fig:Balitsky}. Hence, \eq{Bal1} is not a closed equation that one can solve: instead, it relates different correlators of light-cone Wilson lines to each other. If one tries to write an evolution equation for the new correlator,
\begin{align}
\left\langle \frac{1}{N_c} \, \mbox{tr} \left[ t^b \,  V_{{\vec x}_{1 \perp}} \, t^a \, V^\dagger_{{\vec x}_{0 \perp}} \right] \, U_{{\vec x}_{2 \perp}}^{ba} \right\rangle_Y ,
\end{align}
one would generate correlators with more gluon Wilson lines on that equation's right-hand side, again obtaining an equation which does not close, even if combined with \eq{Bal1}. This way one generates an infinite hierarchy of equations for correlators with progressively higher numbers of Wilson lines. This hierarchy is known as the Balitsky hierarchy \cite{Balitsky:1995ub} and \eq{Bal1} is the first equation in this hierarchy. 
 
Note that one can use 
\begin{equation}
U_{{\vec x}_{\perp}}^{ba} = 2 \, \mbox{tr} \left[ t^a \, V^\dagger_{{\vec x}_{\perp}} \, t^b V_{{\vec x}_{\perp}} \right] 
\end{equation} 
along with the Fierz identity to rewrite \eq{Bal1} as
\begin{align}\label{Bal11}
\partial_Y & \left\langle \frac{1}{N_c} \, \mbox{tr} \left[ V_{{\vec x}_{1 \perp}} \, V^\dagger_{{\vec x}_{0 \perp}} \right]  \right\rangle_Y  = \frac{\as \, N_c}{2 \, \pi^2} \int \, d^2
  x_2 \, \frac{x^2_{10}}{x^2_{20}\,x^2_{21}} \\ & \times \, \left[ \left\langle \frac{1}{N_c} \, \mbox{tr} \left[ V_{{\vec x}_{1 \perp}} \, V^\dagger_{{\vec x}_{2 \perp}} \right] \,  \frac{1}{N_c} \, \mbox{tr} \left[ V_{{\vec x}_{2 \perp}} \, V^\dagger_{{\vec x}_{0 \perp}} \right] \right\rangle_Y - \left\langle \frac{1}{N_c} \, \mbox{tr} \left[ V_{{\vec x}_{1 \perp}} \, V^\dagger_{{\vec x}_{0 \perp}} \right]  \right\rangle \right]_Y . \notag
\end{align}
Even in this form, this is not a closed equation. At large $N_c$ and for large nuclear target one writes
\begin{align}
& \left\langle \mbox{tr} \left[ V_{{\vec x}_{1 \perp}} \, V^\dagger_{{\vec x}_{2 \perp}} \right] \, \mbox{tr} \left[ V_{{\vec x}_{2 \perp}} \, V^\dagger_{{\vec x}_{0 \perp}} \right] \right\rangle_Y \\ & = \left\langle \mbox{tr} \left[ V_{{\vec x}_{1 \perp}} \, V^\dagger_{{\vec x}_{2 \perp}} \right] \right\rangle_Y  \left\langle  \mbox{tr} \left[ V_{{\vec x}_{2 \perp}} \, V^\dagger_{{\vec x}_{0 \perp}} \right] \right\rangle_Y + {\cal O} \left( \frac{1}{N_c^2}, \frac{1}{A^{1/3}} \right) \notag
\end{align} 
and the BK equation \eqref{Sevol} is recovered. Outside the large-$N_c$ limit the equation \eqref{Bal11} is still hard to simplify. 
 
It appears that to generalize the BK equation to the case of all $N_c$ one needs to solve an infinite hierarchy of coupled integro-differential equations. The degrees of freedom in this hierarchy are the fundamental and adjoint infinite light-cone Wilson lines $V_{{\vec x}_\perp}$ and $U_{{\vec x}_\perp}$. They replace the color dipoles as the right degrees of freedom for an arbitrary $N_c$. 

An alternative way to generate any-$N_c$ small-$x$ evolution is to think of energy dependence of the Wilson line correlators as coming from some rapidity-dependent weight functional $W_Y [\alpha]$ needed to obtain expectation values of operators by using the functional averaging
\begin{align}\label{ave_def_aY}
  \langle {\hat O}_\alpha \rangle_Y = \int {\cal D} \alpha \ {\hat
    O}_\alpha \, W_Y [\alpha]
\end{align}
with ${\hat O}_\alpha$ an arbitrary operator. Here
\begin{align}
  \alpha (x^-, {\vec x}_\perp) \, \equiv \, A^+ (x^+=0, x^-, {\vec
    x}_\perp)
\end{align}
and the functional is normalized to one,
\begin{align}
  \int {\cal D} \alpha \, W_Y [\alpha] = 1.
\end{align}

The rapidity dependence of the functional $W_Y [\alpha]$ needs to be obtained from an evolution equation, which one has to construct for it. Following \cite{Mueller:2001uk} one can take some test operator ${\hat O}_\alpha$ made out of infinite light-cone Wilson lines (say, ${\hat O}_{{\vec x}_{1\perp}, {\vec x}_{0\perp}} = V_{{\vec x}_{1\perp}} \otimes V^\dagger_{{\vec x}_{0\perp}}$, a product of two Wilson lines without a color trace) and construct its evolution, obtaining an equation of the type
\begin{align}\label{oev}
  \partial_Y \, \langle {\hat O}_\alpha \rangle_Y = \langle {\cal
    K}_\alpha \otimes {\hat O}_\alpha \rangle_Y = \int {\cal D} \alpha
  \, \left[ {\cal K}_\alpha \otimes {\hat O}_\alpha \right] \, W_Y
  [\alpha],
\end{align}
with some kernel ${\cal K}_\alpha$, which is a differential in $\alpha$ operator whose action on the operator ${\hat O}_\alpha$ is denoted by $\otimes$. At the same time, differentiating \eq{ave_def_aY} with respect to $Y$ we get
\begin{align}\label{oev2}
  \partial_Y \, \langle {\hat O}_\alpha \rangle_Y = \int {\cal D}
  \alpha \ {\hat O}_\alpha \ \partial_Y \, W_Y [\alpha].
\end{align}
Integrating \eq{oev} by parts on its right-hand side we arrive at the evolution equation for $W_Y [\alpha]$, 
\begin{align}\label{JIMWLK}
\partial_Y \, W_Y [\alpha] = {\cal K}_\alpha \otimes W_Y [\alpha].
\end{align}
(We assume that the form of the kernel ${\cal K}_\alpha$ does not change in the integration by parts, which is indeed the case.)

An explicit calculation of the kernel yields (for details see e.g. \cite{Mueller:2001uk,Kovchegov:2012mbw})
\begin{align}
\label{EQ:JIMWLK_Kernel}
\mathcal{K}_{\alpha} = \frac{\alpha_s}{\pi^2}\int d^2 x_\perp \, d^2 y_\perp \, d^2 w_\perp \, & 
\frac{({\vec x}_\perp-{\vec w}_\perp)\cdot({\vec y}_\perp-{\vec w}_\perp)}{|{\vec x}_\perp-{\vec w}_\perp|^2\ |{\vec y}_\perp-{\vec w}_\perp|^2} \left( U_{{\vec w}_\perp} - \frac{U_{{\vec x}_\perp} + U_{{\vec y}_\perp}}{2}\right)^{ba}  \notag \\ & \times \, 
\frac{(ig)^{-2}  \ \delta^2}{\delta\,\alpha^a(x^-<0,{\vec x}_\perp)\ \delta\,\alpha^b(y^->0,{\vec y}_\perp)} .
\end{align}
(This kernel is given in the form recently written in \cite{Cougoulic:2019aja}.) Equation \eqref{JIMWLK} with the kernel from \eq{EQ:JIMWLK_Kernel} is the Jalilian-Marian--Iancu--McLerran--Weigert--Leonidov--Kovner
(JIMWLK)
\cite{Jalilian-Marian:1997dw,Jalilian-Marian:1997gr,Weigert:2000gi,Iancu:2001ad,Iancu:2000hn,Ferreiro:2001qy}
functional evolution equation. The initial condition for JIMWLK evolution is given by the Gaussian functional of the MV model. Since JIMWLK is a functional differential equation, its solution is hard to find analytically, but is possible to construct numerically \cite{Weigert:2000gi,Rummukainen:2003ns,Schlichting:2014ipa}.


\subsection{Unitarization of the BFKL Evolution and Saturation}

No analytic solution of the BK evolution equation \eqref{eqN2} for all dipole sizes and rapidities exists at this time. Instead, there exit approximate analytic solutions which apply in various subset of the $(x_{10}, b_\perp,  Y)$ phase space for $N({\vec x}_{10}, {\vec b}_\perp, Y)$. There are also numerical solutions. A pedagogical presentation of the analytical attempts to solve the BK equation, along with a subset of numerical solutions, can be found in \cite{Kovchegov:2012mbw}. 

Here we want to convey the essential physical properties of the BK solution. To do so, let us consider the following toy version of the BK equation \eqref{eqN2}, 
\begin{align}\label{toyBK}
  \partial_Y N \, = \, \as \, N - \as \, N^2
\end{align}
with $N(Y=0) = N_0 \ll 1$ as the initial condition. In arriving at \eq{toyBK} we have essentially discarded the integral kernel of the full BK equation \eqref{eqN2}, suppressing the transverse coordinate dependence in $N$. The solution of \eq{toyBK} is straightforward to find. It reads
\begin{align}\label{toyBKsol}
N(Y) = \frac{N_0 \, e^{\as \, Y}}{1 + N_0 \, (e^{\as \, Y} - 1)}. 
\end{align}

We see that at moderately large rapidity, when $\as Y \gtrsim 1$, the equation \eqref{toyBKsol} can be linearized in $N_0$, yielding
\begin{align}
N (Y = \ln (1/x)) \approx N_0 \, e^{\as \, Y} \propto e^{\as \, Y} \propto \left( \frac{1}{x} \right)^{\as} .
\end{align}
Similar power-law growth of the amplitudes and cross sections follows from the BFKL equation. Indeed, linearizing \eq{eqN2} by discarding the quadratic in $N$ term on the right, and solving the resulting linear (dipole BFKL) equation, one obtains
\begin{align}\label{N_BFKL}
N ({\vec x}_{10}, {\vec b}_\perp, Y= \ln (1/x)) \propto \left( \frac{1}{x} \right)^{\alpha_P - 1}
\end{align}
with
\begin{align}\label{alphaP}
\alpha_P - 1 = \frac{4 \as N_c}{\pi} \, \ln 2. 
\end{align}
This is the standard solution of the leading-order BFKL equation, with \eq{alphaP} giving the so-called BFKL pomeron intercept. Employing \eq{forw_amp} we see that \eq{N_BFKL} leads to 
\begin{align}
  \sigma_{tot}^{q {\bar q} A} ({\vec x}_\perp, Y= \ln (1/x)) \propto \left( \frac{1}{x} \right)^{\alpha_P - 1} \propto s^{\alpha_P - 1}. 
\end{align}
Clearly, such a power-law growth of the cross section with the center-of-mass energy squared would violate both the black disk limit \eqref{bdl} and the Froissart-Martin bound \eqref{FM} at a sufficiently high $s$. This violation of unitarity has been a problem for BFKL evolution: while the BFKL equation was a great breakthrough in our understanding of high-energy QCD, it was also clear that the equation violated unitarity, and, therefore, something else had to come in at higher energies and modify BFKL equation to make it preserve unitarity. 

Let us return to the full solution \eqref{toyBKsol} of the toy BK equation \eqref{toyBK}. We see that $N(Y)$ given by this solution preserves unitarity. That is, $N(Y) \le 1$ for all values of $Y$. Moreover, $N(Y)$ asymptotically approaches 1 at very large rapidities,
\begin{align}
N(Y) \to 1 \ \ \ \mbox{as} \ \ \ Y \to \infty. 
\end{align}
We see that the toy BK evolution preserves the black-disk limit. Moreover, the dipole scattering amplitude approaches the black disk limit for very large $Y$, corresponding to very high energies. Thus, our toy BK equation \eqref{toyBK} does not have the unitarity violation problem that BFKL equation had. 

\begin{figure}
\begin{center}
\includegraphics[width= 0.6 \textwidth]{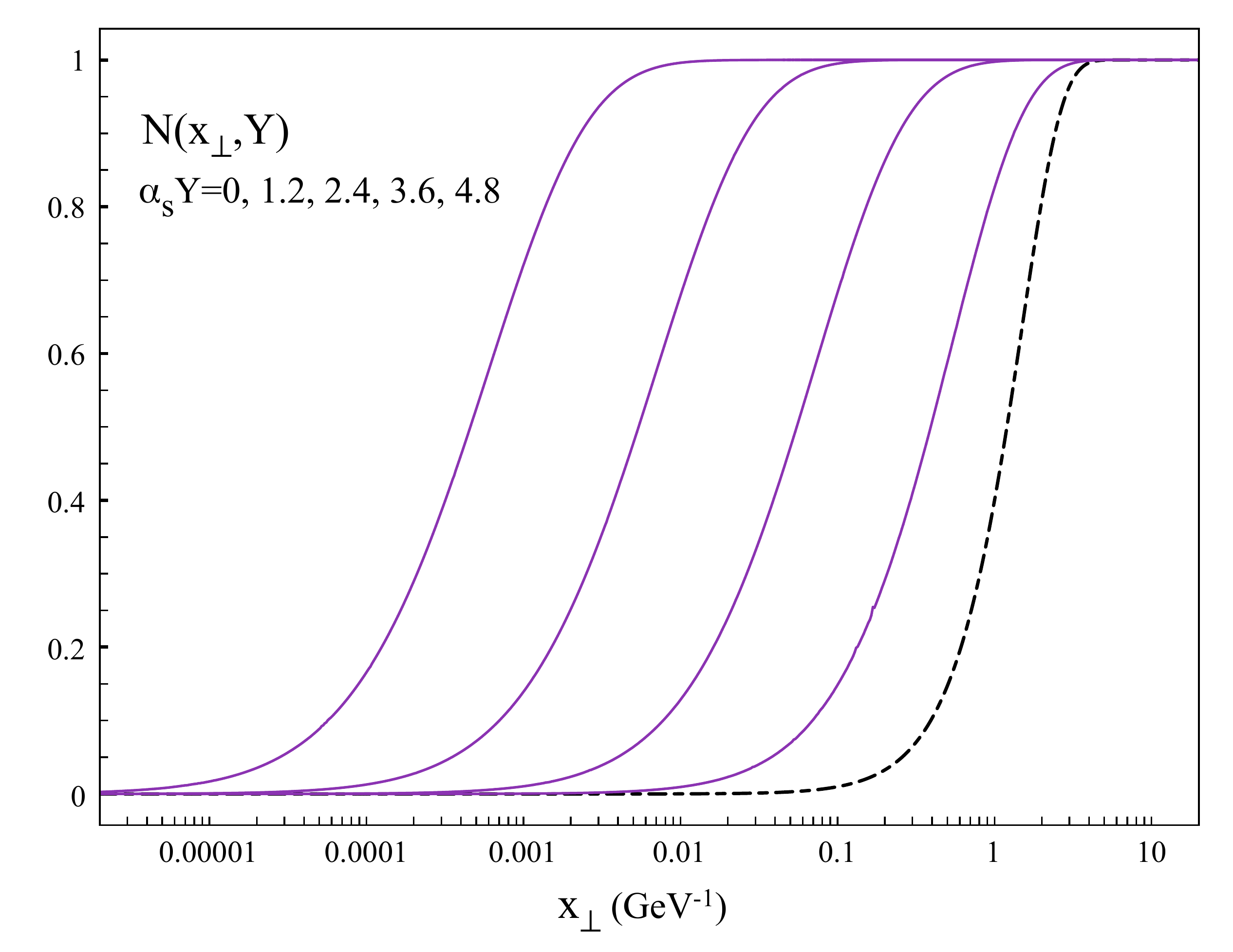} 
\caption{Solution of the BK equation plotted as a function of the dipole size $x_\perp$ for several values of $\as Y$ specified in the legend. The GGM-type initial condition is given by the dashed line.}
\label{fig:BKsol}
\end{center}
\end{figure}

The same conclusion can be applied to the full BK equation \eqref{eqN2}. In \fig{fig:BKsol} we plot a numerical solution of the leading-order BK equation versus the dipole size $x_\perp$. This plot and the underlying solution were constructed by Javier Albacete for \cite{Kovchegov:2012mbw} (see also \cite{Albacete:2004gw,Albacete:2007yr,Albacete:2009fh,Albacete:2010sy}). The $b$-dependence has been suppressed in the solution, such that $N(x_\perp, b_\perp, Y) \approx N (x_\perp, Y)$, which is valid for a large nucleus target. The initial condition for the evolution corresponding to $Y=0$ is given by a GGM-type formula and is plotted by the dashed line in \fig{fig:BKsol} (cf. \fig{fig:saturation} above). The solid curves are given by the BK solution at four different values of $\as Y$ listed in the legend of \fig{fig:BKsol}. There are two ways of thinking about this plot: first of all, one can fix $x_\perp$, and see that with increasing $Y$ the amplitude $N$ raises as well, but never exceeds 1, in qualitative agreement with our toy model \eqref{toyBKsol}. Alternatively, one can think of evolution in \fig{fig:BKsol} moving the initial condition (dashed) curve to the left, in the process somewhat modifying its shape. Again, one always has $N \le 1$ and the black disk limit is not violated. We see that the BK equation leads to unitarization of the BFKL evolution, resolving the problem of unitarity violation by the latter.

\begin{figure}
\begin{center}
\includegraphics[width= 0.6 \textwidth]{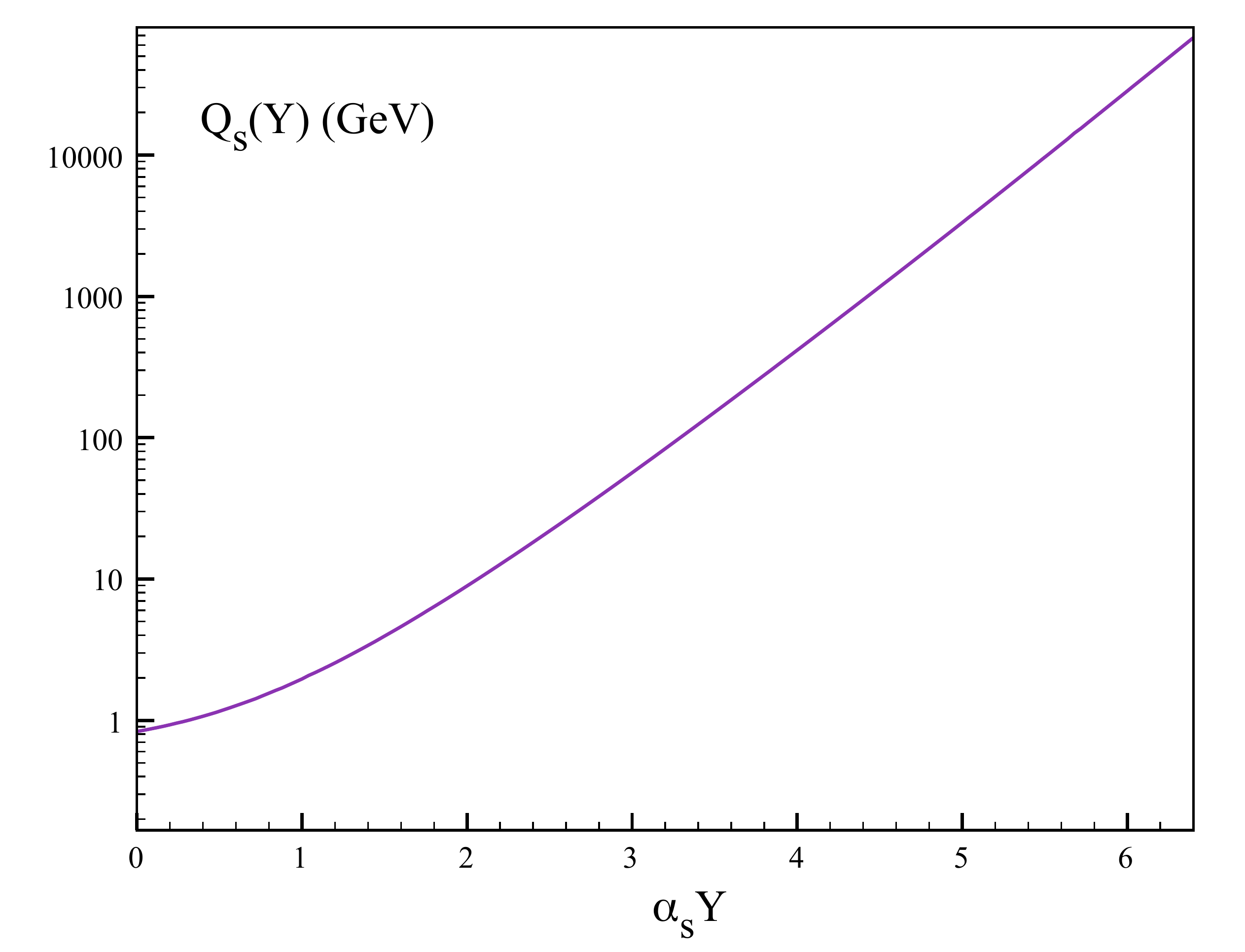} 
\caption{Saturation scale as a function of $\as \, Y$ given by the numerical solution of the BK equation from \fig{fig:BKsol}.}
\label{fig:Qs}
\end{center}
\end{figure}

Let us define the saturation scale $Q_s (Y)$ like in \fig{fig:saturation} as a transition point between the region where $N (x_\perp, Y)$ grows with $x_\perp$ and the region where this growth is tamed, that is, by requiring that $N (x_\perp = 1/Q_s (Y), Y) =$~const, with some order-one constant. From \fig{fig:BKsol} one can clearly see that $1/Q_s (Y)$ decreases with $\as \, Y$. This is why the curves in \fig{fig:BKsol} look like they are moving to the left as $Y$ increases. Therefore, $Q_s (Y)$ increases with $\as \, Y$. A more detailed analytic calculation leads to \cite{Gribov:1984tu,Iancu:2002tr,Mueller:2002zm}
\begin{align}\label{QsY}
Q_s (Y) \approx Q_{s} (0) \, e^{2.44 \frac{\as N_c}{\pi} \, Y}, 
\end{align}
where $Q_{s} (0)$ is the GGM saturation scale \eqref{qsmv}. The saturation scale extracted by Albacete from the numerical solution of the BK equation is shown in \fig{fig:Qs}, and also exhibits growth with rapidity $Y$. 

We see that the non-linear small-$x$ evolution \eqref{eqN2} makes the saturation scale larger, driving it deeper into the perturbative region. Combining Eqs.~\eqref{qsmv} and \eqref{QsY} we see that 
\begin{align}
Q_s^2 (x) \propto A^{1/3} \, \left( \frac{1}{x} \right)^{4.88 \frac{\as N_c}{\pi}}.
\end{align}
We conclude that there are two ways of obtaining a large saturation scale, justifying perturbative approach to small-$x$ physics: one can either use a large nucleus as a target, thus increasing the atomic number $A$. Alternatively (or simultaneously) one can increase the center-of-mass energy, decreasing $x$ and driving $Q_s$ up. The saturation region, defined by $1/Q_s(Y) < x_\perp < 1/\Lambda$, also increases and becomes more prominent with increasing $A$ and decreasing $x$. 

\begin{figure}
\begin{center}
\includegraphics[width= 0.6 \textwidth]{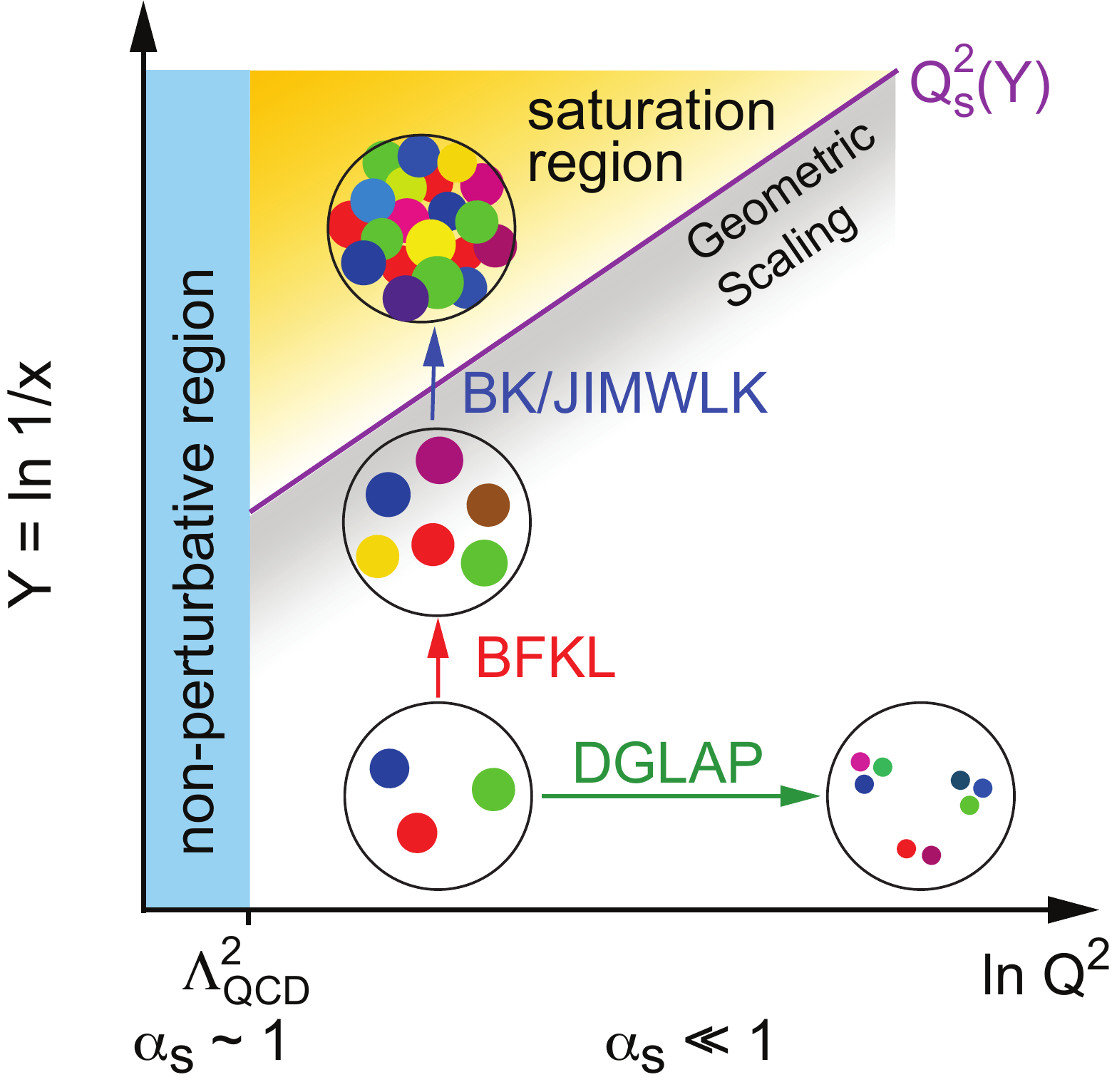} 
\caption{High energy QCD summarized in the $(\ln Q^2, \ln (1/x))$ plane.}
\label{fig:sat_map}
\end{center}
\end{figure}

We summarize the above discussion by the plot in \fig{fig:sat_map}, depicting various regimes of high-energy QCD in the $(\ln Q^2, \ln (1/x))$ plane. First of all we exclude the non-perturbative region with $Q^2 \lesssim \Lambda_{QCD}^2$ as the region where our perturbative methods do not apply. We concentrate on the $Q^2 > \Lambda_{QCD}^2$ perturbative QCD region where $\as \ll 1$. If one is interested in evolution in $Q^2$, it is accomplished by the famous Dokshitzer-Gribov-Lipatov-Altarelli-Parisi (DGLAP) evolution equation \cite{Gribov:1972ri,Altarelli:1977zs,Dokshitzer:1977sg}. Description of this equation is outside our scope here but, for completeness, it is shown in \fig{fig:sat_map}; this is another important equation which bears Lev Lipatov's name.

To study the high energy asymptotics of QCD one needs small-$x$ evolution equations. The BFKL equation allows one to evolve the dipole amplitude or gluon distribution function toward small-$x$. However, at very small $x$, near the saturation region, it has to be replaced by the non-linear BK and JIMWLK evolution equations, as shown in \fig{fig:sat_map}. These latter equations are valid both inside and outside of the saturation region, describing non-linear QCD interactions at high energy. 

We refer the readers interested in the phenomenology of saturation physics to the reviews \cite{Gribov:1984tu,Iancu:2003xm,Weigert:2005us,Jalilian-Marian:2005jf,Gelis:2010nm,Albacete:2014fwa} and the book \cite{Kovchegov:2012mbw}. Let us note that a significant part of the physics program of the Electron-Ion Collider (EIC) \cite{Accardi:2012qut} proposed in the US will be dedicated to the search for and study of the saturation physics.


\section{Spin at Small $x$}

Now let us apply the above formalism to the proton spin problem. For several decades it has been known that the measured spin of quarks and gluons inside the proton does not add up to 1/2 required by helicity sum rules \cite{Jaffe:1989jz,Ji:1996ek,Ji:2012sj}. This is known as the proton spin puzzle or the proton spin crisis (see \cite{Accardi:2012qut,Aschenauer:2016our} and references therein for the current status of the spin puzzle). 

One of the two commonly used helicity sum rules is the Jaffe-Manohar sum rule \cite{Jaffe:1989jz}
\begin{align}
  \label{eq:sum_rule}
  S_q + L_q + S_G + L_G = \frac{1}{2}.
\end{align}
It contains the spin contributions carried by quarks and gluons in the proton
\begin{align}
  \label{eq:net_spin}
  S_q (Q^2) = \frac{1}{2} \, \int\limits_0^1 dx \, \Delta \Sigma (x,
  Q^2), \ \ \ S_G (Q^2) = \int\limits_0^1 dx \, \Delta G (x, Q^2),
\end{align}
along with the quark and gluon orbital angular momenta $L_q$ and $L_G$. Here we have employed helicity parton distribution functions (hPDFs)
\begin{align}
  \label{eq:hPDFs}
  \Delta f (x, Q^2) \equiv f^+ (x, Q^2) - f^- (x, Q^2)
\end{align}
where $f^+$ ($f^-$) denote the number density of partons
with the same (opposite) helicity as the proton, and $f = u, {\bar u},
d, {\bar d}, \ldots , G$. In addition,
\begin{align}
  \label{eq:Sigma}
  \Delta \Sigma (x, Q^2) = \left[ \Delta u + \Delta {\bar u} + \Delta
    d + \Delta {\bar d} + \ldots \right] \! (x, Q^2) . 
\end{align} 

Since no experiment can measure helicity PDFs down to $x=0$, any given experiment can measure the integrands in \eq{eq:net_spin} only down to some $x_{min}$ determined by the experiment's acceptance. Theory input appears to be needed to assess the amount of spin carried by quarks and gluons with $x \in [0, x_{min}]$. Thus, the small-$x$ asymptotics of helicity PDFs needs to be under theoretical control in order to resolve the proton spin puzzle. 

Below we will construct the small-$x$ evolution equations which determine the small-$x$ asymptotics of $\Delta \Sigma$ and $\Delta G$. They are different from the BFKL, BK and JIMWLK evolution equations in several aspects. First of all, at small $x$ the helicity distributions are sub-eikonal: they are suppressed by a power of $x$ as compared to the unpolarized observables we explored above. Another difference, partially compensating for this suppression, is that the resummation parameter of the small-$x$ evolution for helicity is double-logarithmic: the parameter is $\as \, \ln^2 (1/x)$, instead of $\as \, \ln (1/x)$ resummed by the BFKL, BK and JIMWLK equations. 
 
Lev Lipatov was the first to explore resummation of such double-logarithmic parameter,  $\as \, \ln^2 (1/x)$, in QCD. In \cite{Kirschner:1983di}, Kirschner and Lipatov were the first to resum double logarithms of energy for the flavor non-singlet structure function dominated by the QCD Reggeon exchange (see also \cite{Gorshkov:1966ht} for similar resummation in QED). The resummation was accomplished using the so-called infrared evolution equations (IREE), which are a technique different from the one used in the original derivation of the BFKL equation and from the shock-wave formalism we described above. While Lipatov himself never applied the IREE technique to the evolution of helicity distributions, it was done by Bartels, Ermolaev and Ryskin in \cite{Bartels:1995iu,Bartels:1996wc}. Unfortunately, the small-$x$ asymptotics of $\Delta \Sigma$ and $\Delta G$ we construct below using the shock wave approach differs from that found in \cite{Bartels:1996wc} by using IREE. Understanding the origin of this discrepancy is an open problem in the field at the time of writing. 

The details of the calculations leading to the results we present below can be found in \cite{Kovchegov:2015pbl,Kovchegov:2016zex,Kovchegov:2016weo,Kovchegov:2017jxc,Kovchegov:2017lsr,Kovchegov:2018znm,Cougoulic:2019aja}.


\subsection{Polarized ``Wilson lines" and dipoles}

Our analysis here will be parallel to the one above for the unpolarized BK and JIMWLK evolution. As one can show, at small $x$ the quark helicity PDF can be written as \cite{Kovchegov:2015pbl,Kovchegov:2018znm}
\begin{align}\label{DSigma}
 \Delta \Sigma (x, Q^2 )  =  \frac{N_c \, N_f}{2 \pi^3} \, \int\limits_{\Lambda^2/s}^1 \frac{d z}{z}  \, \int\limits_\frac{1}{z \, s}^\frac{1}{z \, Q^2} \, \frac{d x_{10}^2}{x_{10}^2} \, d^2 b_\perp \, G_{10} (z s),
\end{align}
where $N_f$ is the number of quark flavors, $\Lambda$ is an infrared (IR) cutoff, and the center of mass energy squared is $s \approx Q^2 /x$. As before, the impact parameter is defined by ${\vec b}_\perp = ({\vec x}_{1 \perp} + {\vec x}_{0 \perp})/2$.

\begin{figure}
\begin{center}
\includegraphics[width= 0.75 \textwidth]{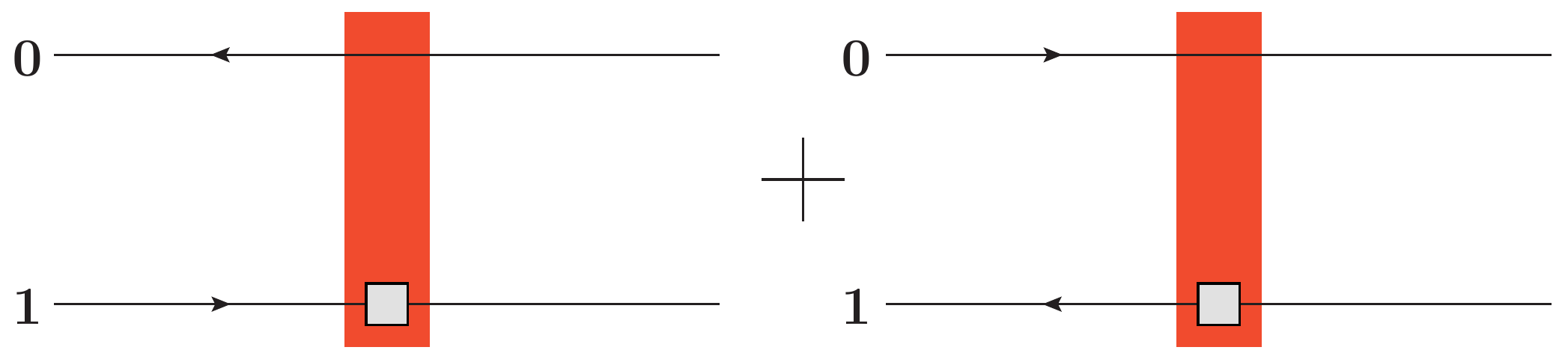} 
\caption{Diagrammatic illustration of the polarized dipole amplitude defined in \eq{eq:Gdef}.}
\label{fig:polarized_dipole}
\end{center}
\end{figure}

Equation~\eqref{DSigma} is the helicity analogue of Eqs.~\eqref{F12sig} and \eqref{forw_amp} above. It relates a physical observable of interest, the quark helicity distribution, to the amplitude of a quark--anti-quark dipole scattering on a target. However, in the case of \eq{DSigma}, the dipole-target scattering has to depend on helicity. More precisely, we are interested in the term in the dipole-target forward scattering amplitude proportional to the product of the projectile and target helicities. The eikonal scattering considered above is independent of helicity. Hence, to obtain a non-zero contribution to \eq{DSigma} we have to introduce sub-eikonal interactions between the dipole and the longitudinally polarized target. This is formally accomplished by defining the so-called polarized dipole scattering amplitude $G_{10} (z s)$ as
\begin{align} 
  \label{eq:Gdef} 
  & G_{10} (z s = \min\{z_1, z_0\} s) \nonumber \\ &\equiv \frac{1}{2 N_c} \,
 \Big\langle \!\! \Big\langle 
  \mbox{T} \, \mbox{tr} \left[V_{{\vec x}_{1 \perp}}^{pol} V_{{\vec x}_{0 \perp}}^\dagger \right] + \mbox{T} \, \mbox{tr} \left[ V_{{\vec x}_{0 \perp}} V_{{\vec x}_{1 \perp}}^{pol \, \dagger} \right] \Big\rangle \!\! \Big\rangle (\min\{z_1, z_0\} s)
\nonumber \\ &\equiv
\frac{z_1 s}{2 N_c} \, \Big\langle  \mbox{T} \, \mbox{tr} \left[V_{{\vec x}_{1 \perp}}^{pol} V_{{\vec x}_{0 \perp}}^\dagger \right] + \mbox{T} \, \mbox{tr} \left[ V_{{\vec x}_{0 \perp}} V_{{\vec x}_{1 \perp}}^{pol
      \, \dagger} \right] \Big\rangle ,
\end{align}
where the double-angle brackets are defined to scale out the
center-of-mass energy $z_1 s$ between the polarized (anti-)quark and the
target. Here the quark/anti-quark 1 carries momentum fraction $z_1$ of some projectile's light-cone (minus) momentum, while the anti-quark/quark 0 carries momentum fraction $z_0$. As usual, T denotes time-ordering. The polarized dipole amplitude is illustrated in \fig{fig:polarized_dipole}. In the polarization-dependent scattering, one (anti-)quark line in the dipole carries polarization information. This is line 1 in our notation. This line interacts with the target in the polarization-dependent sub-eikonal way, which we describe by the so-called ``polarized Wilson line" operator $V_{{\vec x}_{1 \perp}}^{pol}$ in the left panel of \fig{fig:polarized_dipole} and by its conjugate in the right panel. This sub-eikonal helicity-dependent interaction is denoted by the box in \fig{fig:polarized_dipole}; its structure will be clarified shortly. The unpolarized line 0 interacts with the shock wave target in the standard eikonal way described by the infinite light-cone Wilson line in \eq{Vline_inf}. 

\begin{figure}[ht]
\begin{center}
\includegraphics[width=  \textwidth]{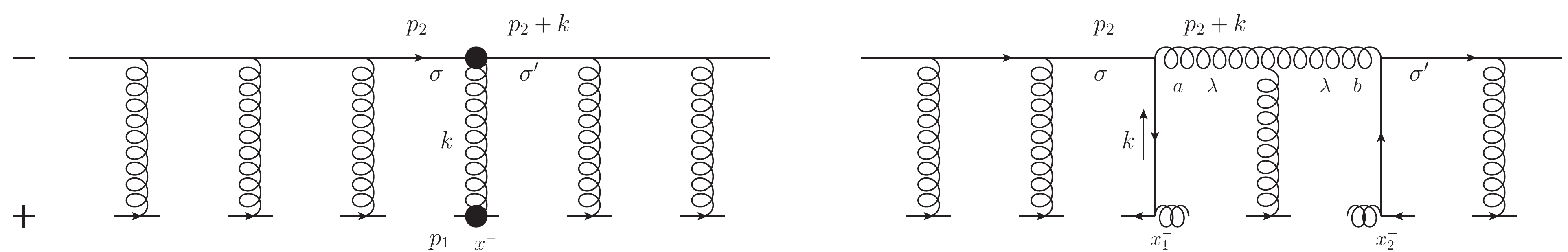} 
\caption{Two contributions to the polarized fundamental Wilson line in a background field. Filled circles at the quark-gluon vertices
  denote the spin-dependent sub-eikonal scattering.}
\label{vpol}
\end{center}
\end{figure}

To determine the form of the ``polarized Wilson line" operator $V_{{\vec x}_{1 \perp}}^{pol}$ we again use the analogue to the unpolarized case and consider scattering of a quark on a GGM type of a target, say a large nucleus with many nucleons. The difference now is that one of the nucleons has to be longitudinally polarized, while the interaction with this nucleon should be helicity-dependent, and, hence, sub-eikonal. The scattering amplitude giving the helicity-dependent contribution to the quark scattering on a shock wave target is shown in \fig{vpol}, where the black circles in the left panel denote sub-eikonal helicity-dependent parts of the quark-gluon interactions. 

A detailed calculation of the ``polarized Wilson line" operator in the fundamental representation can be found in \cite{Kovchegov:2017lsr,Kovchegov:2018znm} (see also \cite{Chirilli:2018kkw}) and yields
\begin{align} 
  \label{eq:Wpol_all} 
  & V^{pol}_{{\vec x}_\perp} = \frac{i g p_1^+}{s} \,
  \int\limits_{-\infty}^\infty d x^- \, V_{{\vec x}_\perp} [+\infty, x^-] \: 
  F^{12} (x^-, {\vec x}_\perp) \: V_{{\vec x}_\perp} [x^- , -\infty]  - \frac{g^2 p_1^+}{s} \!\!
  \int\limits_{-\infty}^\infty d x_1^- \!\! \int\limits_{x_1^-}^\infty d x_2^- \notag  \\ & \times V_{{\vec x}_\perp} [+\infty, x_2^-] \:  t^b \, {\psi}_\beta (x_2^-, {\vec x}_\perp)  \, U_{{\vec x}_\perp}^{ba} [ x_2^-,  x_1^-] \left[ \frac{1}{2} \, \gamma^+ \, \gamma^5 \right]_{\alpha\beta} \!\! {\bar \psi}_\alpha (x_1^-, {{\vec x}_\perp}) \, t^a  \: V_{{\vec x}_\perp} [x_1^- , -\infty]. 
\end{align}
While the target still generates the eikonal gluon field $A^+$ contributing to the Wilson lines in \eq{eq:Wpol_all}, at the sub-eikonal level it also generates a (helicity-dependent part of the) transverse gluon field ${\vec A}_\perp$ which gives the 12 component of the gluon field strength tensor $F^{12}$ in the first term of  \eq{eq:Wpol_all}. In addition, the target generates the (sub-eikonal) quark and anti-quark fields $\psi_\beta$ and ${\bar \psi}_\alpha$ with $\alpha, \beta$ the Dirac spinor indices: these quark fields lead to the second term in \eq{eq:Wpol_all}. Here $p_1^+$ is the large momentum component of the target nucleons (see \fig{vpol}). 

A similar calculation for a longitudinally polarized gluon scattering on a polarized target leads to the adjoint ``polarized Wilson line" operator \cite{Kovchegov:2018znm}
\begin{align} 
  \label{M:UpolFull}
  & (U_{{\vec x}_\perp}^{pol})^{ab} = \frac{2 i \, g \, p_1^+}{s}
  \int\limits_{-\infty}^{+\infty} dx^- \: \left( U_{{\vec x}_\perp}[+\infty, x^-] \:
  {\cal F}^{12} (x^+ =0 , x^- , {{\vec x}_\perp}) \: U_{{\vec x}_\perp} [x^- , -\infty] \right)^{ab} \notag \\ &  - \frac{g^2 \, p_1^+}{s} \, \int\limits_{-\infty}^\infty d x_1^- \, \int\limits_{x_1^-}^\infty d x_2^- \, U^{aa'}_{{\vec x}_\perp} [+\infty, x_2^-] \,  {\bar \psi} (x_2^-, {{\vec x}_\perp}) \, t^{a'} \, V_{{\vec x}_\perp} [x_2^-, x_1^-] \, \frac{1}{2} \, \gamma^+ \gamma_5 \, t^{b'} \notag \\ & \times \,  \psi (x_1^-, {{\vec x}_\perp}) \, U^{b'b}_{{\vec x}_\perp} [x_1^-, -\infty] - \mbox{c.c.}  . 
\end{align}
Now ${\cal F}^{12}$ is the gluon field strength tensor in the adjoint representation. 

Apart from some formal complexity, the overall picture of helicity-dependent scattering is very similar to the unpolarized case considered above. We need to construct and solve the evolution equation for the polarized dipole amplitude $G_{10} (z s)$. Equation~\eqref{DSigma} would then give us the flavor-singlet quark helicity distribution. Similar calculation, with a slightly different polarized dipole amplitude, would give us the gluon helicity distribution $\Delta G$ at small $x$ \cite{Kovchegov:2017lsr}. For the quark flavor non-singlet helicity PDF, one needs a different polarized dipole definition from \eq{eq:Gdef}, which, in turn, satisfies different evolution equations from what we construct below, see \cite{Kovchegov:2016zex,Kovchegov:2018znm} along with  \cite{Bartels:1995iu}: however, the flavor non-singlet distribution, while an interesting quantity in its own right, is not a part of the helicity sum rules such as the one in \eq{eq:sum_rule} and will not be discussed here.


\subsection{Small-$x$ evolution for polarized dipoles}

\begin{figure}[h]
\centering
\includegraphics[width= 0.95 \textwidth]{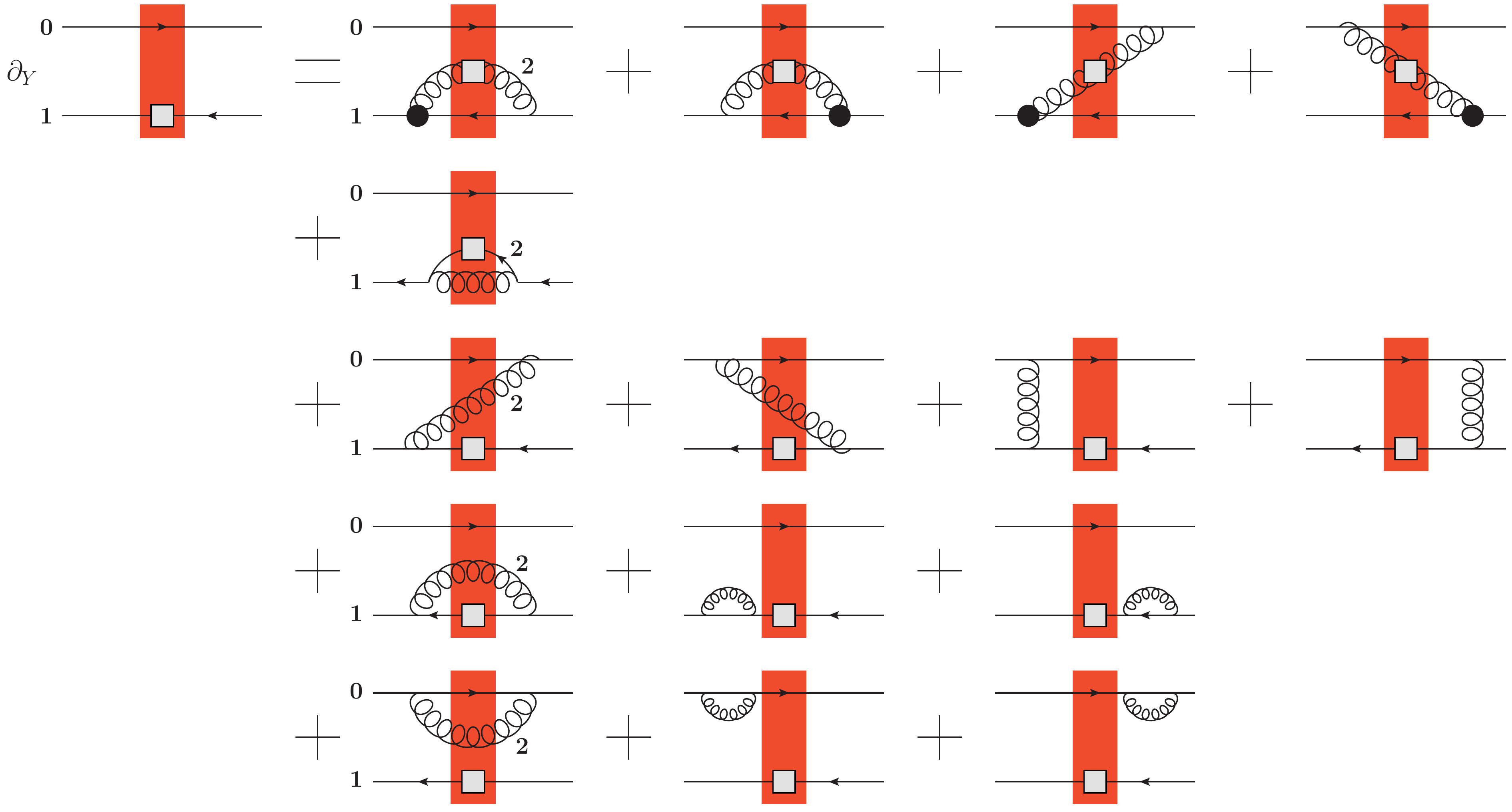}
\caption{One step of the polarized fundamental dipole
  small-$x$ evolution. Black circles
  represent spin-dependent (sub-eikonal) soft gluon emission
  vertices.}
\label{es}
\end{figure}

Our next step is to construct an evolution equation for the polarized dipole amplitude $G_{10} (z s)$ in \eq{eq:Gdef}. In the operator language for the background field method, like that employed in \cite{Balitsky:1995ub} for the derivation of the unpolarized small-$x$ evolution equations, one separates the gluon field into the classical $A^\mu_{cl}$ and quantum $a^\mu$ parts by writing $A^\mu = A^\mu_{cl} + a^\mu$, and integrates out the quantum field keeping only the terms up to the required order in the coupling constant $\as$ (ditto for the quark fields). Equivalently, one can think of this evolution in terms of Feynman diagrams, similar to the unpolarized case discussed above. Because the expression for the evolution of both terms in the polarized dipole amplitude in \fig{fig:polarized_dipole} is somewhat lengthy, we show the (flavor-singlet) evolution only for one of those terms in \fig{es}. In the first row of \fig{es}, shaded circles denote the sub-eikonal helicity-dependent gluon emissions, just like in \fig{vpol}. These vertices originate in the $F^{12}$ contribution in \eq{eq:Wpol_all}. The second row of \fig{es} contains emission of the soft quark, originating in the second term of \eq{eq:Wpol_all}. The remaining rows three to five in \fig{es} correspond to the standard eikonal unpolarized BK/JIMWLK evolution described above. 

The diagrams in \fig{es} lead to the following evolution equation \cite{Kovchegov:2015pbl}:
\begin{align}\label{evol4}
  \frac{1}{N_c} \, \left\langle \!\! \left\langle \mbox{tr} \left[
        V_{{\vec x}_{0 \perp}} \, V_{{\vec x}_{1 \perp}}^{pol \, \dagger} \right]
    \right\rangle \!\! \right\rangle & (z s) = \frac{1}{N_c} \,
  \left\langle \!\! \left\langle \mbox{tr} \left[ V_{{\vec x}_{0 \perp}} \,
        V_{{\vec x}_{1 \perp}}^{pol \, \dagger} \right] \right\rangle \!\!
  \right\rangle_0 (z s) + \frac{\as}{2 \pi} \int\limits_{\Lambda^2/s}^z
  \frac{d z_2}{z_2} \!\! \int\limits_{1/(z_2 s)} \!\!\!\! \frac{d x_{21}^2}{x_{21}^2}
  \notag \\ & \times \Bigg\{ \theta (x_{10} - x_{21}) \, \frac{2}{N_c}
    \left\langle \!\! \left\langle \mbox{tr} \left[ t^b \,
          V_{{\vec x}_{0 \perp}} \, t^a \, V_{{\vec x}_{1 \perp}}^{\dagger}
        \right] \, U^{pol \, ba}_{{\vec x}_{2 \perp}} \right\rangle \!\!
    \right\rangle (z_2) \notag \\ &  + \theta (x_{10}^2 z -
  x_{21}^2 z_2) \, \frac{1}{N_c} \left\langle \!\! \left\langle
      \mbox{tr} \left[ t^b \, V_{{\vec x}_{0 \perp}} \, t^a \,
        V_{{\vec x}_{2 \perp}}^{pol \, \dagger} \right] \, U^{ba}_{{\vec x}_{1 \perp}}
    \right\rangle \!\! \right\rangle (z_2) \notag \\ & + \theta (x_{10}
  - x_{21})  \frac{1}{N_c} \left[ \left\langle \!\! \left\langle
          \mbox{tr} \left[ V_{{\vec x}_{0 \perp}} \, V^{\dagger}_{{\vec x}_{2 \perp}} \right] \, \mbox{tr} \left[
            V_{{\vec x}_{2 \perp}} \, V^{pol \, \dagger}_{{\vec x}_{1 \perp}} \right]
        \right\rangle \!\! \right\rangle (z_2) \right. \notag \\ & \left. - N_c \left\langle \!\!
        \left\langle \mbox{tr} \left[ V_{{\vec x}_{0 \perp}} \,
            V_{{\vec x}_{1 \perp}}^{pol \, \dagger} \right] \right\rangle \!\!
      \right\rangle (z_2) \right] \Bigg\}.
\end{align}
Lines 2 and 3 of \eq{evol4} correspond to rows 1 and 2 in \fig{es} respectively, while lines 4 and 5 in \eq{evol4} are given by the unpolarized small-$x$ evolution in rows 3-5 of \fig{es} (cf. \eq{Bal11}). (Here $z_2$ is the longitudinal momentum fraction carried by the soft parton 2, which can be a gluon or a quark.) The evolution equation \eqref{evol4} is written in the integral form, with the $\left\langle \! \left\langle \ldots \right\rangle \!
  \right\rangle_0$ inhomogeneous term given by the initial conditions for the evolution. (The BK equation can also be written in an integral form \cite{Kovchegov:1999yj}.)

As we mentioned above, there is an important difference between the helicity evolution \eqref{evol4} and the BK/JIMWLK equations. Let us recall the lifetime ordering \eqref{lifetime_ord} of the soft gluon emissions in the unpolarized evolution. The same lifetime ordering applies here to spin evolution. In the unpolarized evolution case, the lifetime ordering was included by simply imposing the longitudinal momentum ordering \eqref{long_ord}, or, equivalently, the longitudinal momentum fraction ordering  
\begin{align}\label{zord}
z_0, z_1 \gg z_2 \gg z_3 \gg \ldots . 
\end{align}
Equations \eqref{lifetime_ord} and \eqref{zord} are equivalent only if the partons' transverse momenta are comparable. (Note that $z_i = k_i^-/p_2^-$, where $p_2^-$ is the large light-cone momentum of the projectile.) In general, the condition \eqref{lifetime_ord} translates into 
\begin{align}\label{lifetime_ord2}
\min \{ z_1, z_0 \} \, x_{10}^2 \gg z_2 \, x_{21}^2 \gg z_3 \, x_{32}^2 \gg \ldots
\end{align}
in the transverse coordinate space. (At this point we have randomly picked the distance $x_{21}$ over $x_{20}$ and picked $x_{32}$ over other transverse distances at that step of evolution.) 

The unpolarized evolution equation \eqref{Bal11} is both the IR and ultra-violet (UV) finite: the expression in the square brackets on the right goes to zero when ${\vec x}_{2\perp} \to {\vec x}_{1\perp}$ and ${\vec x}_{2\perp} \to {\vec x}_{0\perp}$ regulating the divergence in the integral kernel in those limits. The same integral kernel goes to zero reasonably fast when ${\vec x}_{2\perp} \to \infty$, ensuring the IR convergence of the integral. Since the $x_2$ integral in \eq{Bal11} is thus convergent, the typical ``daughter" dipole sizes are not much larger or smaller than the ``parent" dipole size, that is, $x_{21} \approx x_{20} \approx x_{10}$. This way, \eq{zord} automatically leads to \eq{lifetime_ord2}. Now, let us consider the evolution in \eq{evol4}. The $x_2$ integral in it is no longer UV convergent. For instance, analyzing the terms in the last two lines of \eq{evol4} we see that (unlike the unpolarized case) they do not cancel in the ${\vec x}_{2\perp} \to {\vec x}_{1\perp}$ limit. This UV divergence is regularized by the highest energy scale in the problem, the effective center-of-mass energy squared $z_2 s$ between the soft parton and the target: that is why we have the $1/(z_2 s)$ lower limit on the $x_{21}$ integral in \eq{evol4}. Moreover, the divergence in the $x_{21}^2 \to 1/(z_2 s)$ limit is logarithmic: hence, in the case of helicity evolution the transverse integral generates another logarithm of energy. We have simplified the integral kernel in arriving at \eq{evol4} to extract this logarithm while discarding non-logarithmic transverse integral contributions. That is why the $x_{21}$ integral is logarithmic in \eq{evol4}. We conclude that $x_{21} \ll x_{20} \approx x_{10}$ in the last two terms on the right of \eq{evol4}: this condition is enforced by $\theta (x_{10} - x_{21})$ multiplying those terms.  In addition, consider the second term in the curly brackets of \eq{evol4}. The $x_{21}$ integral in its kernel is UV and IR divergent. The IR divergence means that one may have $x_{21} \approx x_{20} \gg x_{10}$, such that the condition \eqref{zord} no longer automatically leads to the condition \eqref{lifetime_ord2}. Hence, the condition \eqref{lifetime_ord2} has to be imposed explicitly in \eq{evol4}. This is what gives rise to $\theta (x_{10}^2 z -
  x_{21}^2 z_2)$ in \eq{evol4}. This theta-function in fact multiplies all terms on the right of \eq{evol4}: however, in some terms it has been replaced by a stronger condition, $\theta (x_{10} - x_{21})$, resulting from the simplification of the integral kernel to extract the second logarithm of energy, as we have just described. 

Let us pose here to describe the resummation parameter. At each step of small-$x$ helicity evolution we emit a soft parton, generating a power of the coupling $\as$. Similar to the unpolarized evolution, the integral over the longitudinal momentum fraction $z_2$ of this parton gives a logarithm of energy, or, equivalently, of Bjorken $x$. In addition, as we have just discussed, and unlike the unpolarized evolution, the transverse momentum/position integral gives another $\ln (1/x)$ for helicity evolution. The resulting resummation parameter for helicity evolution is $\as \, \ln^2 (1/x)$: the powers of this parameter are resummed by the equation \eqref{evol4}. This parameter does not exist for the unpolarized evolution, which resums power of $\as \, \ln (1/x)$ at leading order, which is a smaller parameter at small $x$. Hence, the resummation parameter for helicity evolution is larger than the one for BFKL, BK and JIMWLK equations. At the same time, helicity evolution is sub-eikonal, and helicity distributions are suppressed by an extra power of $x$ as compared to the unpolarized parton distributions at small $x$.  As we mentioned above, Lev Lipatov, together with Kirschner, were the first to discover this double-logarithmic $\as \, \ln^2 (1/x)$ resummation parameter in \cite{Kirschner:1983di}.  

Returning to \eq{evol4} we notice that it is not closed, just like the first equation in Balitsky hierarchy \eqref{Bal11}. Once again, the equation closes if we take the large-$N_c$ limit, and, in addition, in the large-$N_c \& N_f$ limit (with $N_f$ the number of quark flavors) \cite{Kovchegov:2015pbl}. In the large-$N_c$ limit we obtain (adding the contribution of the other term in \fig{fig:polarized_dipole}) \cite{Kovchegov:2015pbl,Kovchegov:2016zex}
\begin{subequations}\label{GNc}
\begin{align}\label{GNc1}
& G_{10} (z s) = G_{10}^{(0)} (z s) + \frac{\alpha_s \, N_c}{2 \pi} \int\limits_{\frac{1}{s \, x_{10}^2}}^{z}
\frac{dz'}{z'} \int\limits^{x_{10}^2}_\frac{1}{z' s} \frac{d x_{21}^2}{x_{21}^2} \: \left[ \Gamma_{10,21} (z' s) + 3 \, G_{21} (z' s)  \right], \\
\label{GNc2}
& \Gamma_{10,21} (z' s) = \Gamma_{10,21}^{(0)} (z' s) + \frac{\alpha_s \, N_c}{2 \pi} \int\limits_{\min \{ \Lambda^2, \frac{1}{x_{10}^2} \} / s }^{z'}
\frac{dz''}{z''} \int\limits^{\min \{ x_{10}^2, x_{21}^2 z'/z'' \} }_\frac{1}{z'' s} \frac{d x_{32}^2}{x_{32}^2} \\ & \hspace*{5cm} \times \left[ \Gamma_{10,32} (z'' s) + 3 \, G_{32} (z'' s)  \right]. \notag
\end{align}
\end{subequations}
Here we had to introduce an auxiliary function $\Gamma$, termed the ``neighbor
dipole amplitude'', in which further evolution is constrained by the
lifetime of an adjacent dipole: the evolution in dipole 10 may depend in the size of the neighbor dipole 21. This happens when the lifetime ordering condition \eqref{lifetime_ord2} for the evolution in dipole 10 is dominated by the lifetime of this neighbor dipole, that is, if the lifetime of dipole 21, $z' x_{21}^2$, is much shorter than the lifetime of dipole 10, $z' x_{10}^2$. The resulting neighbor dipole amplitude depends on ${\vec x}_{10}$, $x_{21}$ and $z'$ and is denoted by $\Gamma_{10,21} (z' s)$ (for more details, see \cite{Kovchegov:2015pbl}). The inhomogeneous terms $G_{10}^{(0)} (z s) = \Gamma_{10,21}^{(0)} (z' s)$ in Eqs.~\eqref{GNc} are determined by the initial conditions for the helicity evolution. 
  
Note that after taking the large-$N_c$ limit of \eq{evol4} we have also put all the unpolarized $S$-matrices \eqref{Sdef} to 1, which corresponds to neglecting saturation effects in the resulting equations \eqref{GNc}. This is done in order to keep only the leading double-logarithmic evolution (powers of $\as \, \ln^2 (1/x)$); saturation corrections appear at the leading-$\ln (1/x)$ level, that is, at order-$\as \, \ln (1/x)$, which is a higher-order correction for helicity evolution. 


\subsection{Small-$x$ asymptotics for quark and gluon helicity distributions}

To determine the small-$x$ asymptotics of the quark helicity distribution one has to solve Eqs.~\eqref{GNc}. 
These equations can be solved numerically
\cite{Kovchegov:2016weo} and analytically \cite{Kovchegov:2017jxc} giving
\begin{align}
  G(x_{10}^2 , z s) \propto  (z s \, x_{10}^2)^{\alpha_h^q} 
\end{align}
with
\begin{align} 
  \label{M:ahel}
  \alpha_h^q = \frac{4}{\sqrt{3}} \sqrt{\frac{\alpha_s N_c}{2\pi}}
  \approx 2.31 \sqrt{\frac{\alpha_s N_c}{2\pi}} .
\end{align}
Employing \eq{DSigma} this result gives us the small-$x$ asymptotics for the quark helicity distribution
\begin{align}\label{dq_final}
  \Delta \Sigma (x, Q^2) \sim \left(
    \frac{1}{x} \right)^{\alpha_h^q}.
\end{align}
Note that this result was derived in the large-$N_c$ limit. At the time of writing, the solution of the large-$N_c \& N_f$ helicity evolution equations from  \cite{Kovchegov:2015pbl,Kovchegov:2018znm} has not been constructed, and the resulting corrections to \eq{dq_final} due to bringing back the quarks into the evolution are not yet known. 

Also note that \eq{dq_final} gives us the behavior of $\Delta \Sigma$ at small $x$, but still outside of the saturation region. While detailed inclusion of saturation effects into equations \eqref{GNc} can only be accomplished when the leading-logarithmic corrections (powers of $\as\, \ln (1/x)$) are summed up consistently, the preliminary conclusion one can draw from the version of equations \eqref{GNc} with the saturation corrections included, as derived in \cite{Kovchegov:2015pbl}, is that $\Delta \Sigma \lesssim$~const inside the saturation region and, therefore, this region contributes very little to the net quark spin in \eq{eq:net_spin}. 

An analysis similar to above can be applied to the gluon helicity distribution \cite{Kovchegov:2017lsr}. Again, in the large-$N_c$ limit and outside the saturation region, one obtains
\begin{align}\label{dG_final}
  \Delta G  (x, Q^2) \sim \left(
    \frac{1}{x} \right)^{\alpha_h^G} \ \ \ \mbox{with} \ \ \ \alpha_h^G =
  \frac{13}{4 \sqrt{3}} \, \sqrt{\frac{\as \, N_c}{2 \pi}}
  \approx 1.88 \, \sqrt{\frac{\as \, N_c}{2 \pi}}.
\end{align}
The small-$x$ asymptotics of $\Delta G$ in the large-$N_c \& N_f$ limit is also not yet known. Similar to the quark helicity distribution, we expect $\Delta G \lesssim$~const in the saturation region. 

Let us note again that the results shown in Eqs.~\eqref{dq_final} and \eqref{dG_final} disagree with the earlier conclusions regarding the small-$x$ asymptotics of helicity distributions reached in \cite{Bartels:1996wc}. This discrepancy is an open problem in the field. A possible resolution of this problem was proposed in \cite{Kovchegov:2016zex}.

Equations \eqref{dq_final} and \eqref{dG_final} are the leading perturbative results for the small-$x$ asymptotics of $\Delta \Sigma$ and $\Delta G$ at large $N_c$ and outside the saturation region. These results can be used to estimate the amount of proton spin carried by the small-$x$ quarks and gluons in the proton. Initial phenomenological analysis has been performed in \cite{Kovchegov:2016weo,Kovchegov:2017lsr}, indicating a potential numerical importance of the small-$x$ region for the proton spin balance. Higher-order corrections to the helicity evolution equations derived above, along with a more detailed phenomenology implementing those corrections, may reduce the theoretical error bars, allowing one to make precise predictions for the helicity PDFs at small $x$ to be measured at EIC. If such predictions are confirmed by the EIC data, one may then gain enough confidence to extrapolate the helicity PDFs down to $x \to 0$, thus providing a robust theoretical assessment of the amount of proton spin at small $x$.


\section{Conclusions}

In this Chapter we have presented two developments initiated by the works of Lev Lipatov. The work on saturation and unitarity started shortly after the derivation of the BFKL equation by Lipatov and collaborators. It would have been impossible without BFKL. These days the field of parton saturation at small-$x$ is a developed field, with a large number of active researchers and an integral component of most QCD-themed conferences and workshops. The search for and discovery of saturation physics is a large part of the physics case for the proposed Electron-Ion Collider in the US \cite{Accardi:2012qut}.  

While Lev Lipatov himself has not worked on the proton spin puzzle, he was the first to study the the double-logarithmic resummation parameter $\as \, \ln^2 (1/x)$ in QCD. Resummation of this parameter is important for the determination of small-$x$ asymptotics of helicity PDFs and may prove to be a crucial tool for resolving the proton spin puzzle. The work on helicity PDFs at small $x$ is only beginning, but has a promising future, especially when the EIC produces new data on longitudinal spin at small $x$.


\section*{Acknowledgments}

I would like to thank Al Mueller, Genya Levin, and Larry McLerran for their collaboration and for teaching me a lot about high energy QCD when I was a junior researcher. I also thank Dan Pitonyak and Matt Sievert for their collaboration on the papers which led to the second part of this chapter. 

This material is based upon work supported by the U.S. Department of
Energy, Office of Science, Office of Nuclear Physics under Award
Number DE-SC0004286.


\end{document}